\documentstyle[12pt]{article}
\catcode`@=11
\def\seceqaa{\@addtoreset{equation}{section}
           \def\theequation{A\arabic{equation}}}
\def\seceqbb{\@addtoreset{equation}{section}
           \def\theequation{B\arabic{equation}}}
\catcode`@=12
\begin{document}
\hfill hep-th/0506224\\
 \begin{center}
{\Large \bf Uplifting the Iwasawa}
\vskip 0.1in
\Large Anne Franzen$^{(i)(ii)}$\footnote{e-mail:anne-franzen@web.de}, Payal
Kaura$^{(ii)}$\footnote{e-mail: pa123dph@iitr.ernet.in}, 
Aalok Misra$^{(ii),(iii)}$\footnote{e-mail: aalokfph@iitr.ernet.in}
and Rajyavardhan Ray$^{(ii)}$\footnote{e-mail: riitrpph@iitr.ernet.in}\\
(i) \Large Department of Physics, 
Rheinisch-Westf\"{a}lische Technische Hochschule Aachen, 52056 Aachen, Germany\\
(ii) Department of Physics, Indian Istitute of Technology Roorkee, 
Roorkee 247 667, India\\
(iii) Jefferson Physical Laboratory, Harvard University, Cambridge, MA 02138,  
USA\\
\vspace{0.5 cm}
\end{center}

\begin{abstract}
The
Iwasawa manifold is uplifted to 
seven-folds of either $G_2$ holonomy or $SU(3)$ structure,
explicit new metrics for the same having been constructed in this work.
We uplift the Iwasawa manifold to a  
$G_2$ manifold through "size" deformation
(of the Iwasawa metric), 
via Hitchin's Flow equations, showing also the impossibility of the uplift
for ``shape" and ``size" deformations (of the Iwasawa metric). 
Using results of \cite{dp}, 
we also uplift the Iwasawa manifold to a 7-fold with $SU(3)$ structure 
through "size" and "shape" deformations via generalisation of Hitchin's Flow 
equations. For seven-folds with $SU(3)$-structure, 
the result could be interpreted as $M5$-branes 
wrapping two-cycles embedded in the seven-fold (as in \cite{dp}) - 
a warped product of either a special hermitian
six-fold or a balanced six-fold with the unit interval.
There can be no uplift to
seven-folds of $SU(3)$ structure involving non-trivial ``size" and ``shape"
deformations (of the Iwasawa metric) retaining the ``standard complex 
structure" - the uplift generically makes one move in the space of almost
complex structures such that one is neither at the 
standard complex structure point nor at the "edge". Using
the results of \cite{confimrs}, we show that given two 
"shape deformation" functions, and the dilaton, one can construct 
a Riemann surface obtained via Weierstra$\ss$ representation for the conformal
immersion of a surface in ${\bf R}^l$, for a suitable $l$, with the condition
of having conformal immersion being a quadric in ${\bf CP}^{l-1}$.
\end{abstract}

\clearpage
\section{Introduction}
 
Unlike the standard model, which contains many free parameters, it is 
expected that string theory, should be able to do a much better job, 
as regards
explaining things like quark and lepton masses and the gauge hierarchy, etc, 
from the vacuum expectation values (vev's) of the moduli fields.
In addition to giving vevs to the moduli (\cite{kachruetal,bbdg}), 
string compactifications in the presence of fluxes, 
can also give a positive cosmological constant \cite{kklt}, making the 
study even more phenomenologically interesting. 
The inclusion of fluxes necessarily requires the introduction of
warp factors in the metric \cite{strominger,bbdg,bbdgs,bd,ck}, 
which can be argued to provide a 
mechanism for generating the large hierarchy of scales - the (not-so)
recent TeV-scale quantum gravity \cite{rs} 
proposal being particularly significant. 
The study of fluxes is also of much importance from the point of view
of open/closed (topological) string dualities \cite{malda,gv}.
Thus, turning on fluxes is of great
importance for establishing connection between string theory and the 
observable universe.

Turning on $NS-NS$ fluxes in, e.g., Heterotic theory,
the internal manifold cannot be K\"{a}hler anymore implying that one then has 
torsion. As shown in \cite{strominger,Cardosoetal}, Iwasawa manifold, an
example of a ``half-flat" manifold (see also \cite{gurrierietal} and
references therein) satisfies
the conditions for ${\cal N}=1$ supersymmetry in the presence of $NS-NS$
fluxes. Based on \cite{Cardosoetal,dp}, we study the explicit uplifts of
the Iwasawa metric at the ``standard complex structure" point of the moduli
space of almost complex structures, to seven-folds with either $G_2$ holonomy
or $SU(3)$ structure. 

The plan of the paper is as follows. In Section 2, we give a brief review
of six-folds with $SU(3)$ structure, the relevant torsion classes 
 and the Iwasawa manifold. In Section 3,
we discuss the uplift of the Iwasawa manifold to seven-folds of $G_2$ holonomy
using Hitchin's flow equations using ``size" deformations and we show
the impossibility of the uplift using ``size" and ``shape" deformations
of the Iwasawa via Hitchin's flow equations. We then give a brief 
review of seven-folds with $SU(3)$ structure and the relevant torsion
classes and then discuss the uplift of the Iwasawa (at the 
``standard complex structure" point) to seven-folds with $SU(3)$ structure
using ``size" and ``shape" deformations via generalizations of the Hitchin's
flow equations. Section 4 has the interpretation of the non-vanishing
Bianchi identity for $M$-theory compactified on seven-folds with 
$SU(3)$ structure, in the presence of $G$-flux sourced by 
 $M5$-branes wrapping two-cycles embedded in the seven-fold 
(analagous to \cite{dp}) and the generic
movement in the moduli space of almost complex structures on the Iwasawa
by ``size" and ``shape" deformations of the Iwasawa metric.
Using the results of \cite{confimrs}, we show that one can associate a 
Riemann surface with a pair of ``shape deformation" functions and the dilaton,
via the Weierstra$\ss$ representation for the conformal immersion of a surface
in ${\bf R}^l$ for a suitable $l$. 
Section 5 has the conclusion. There are two appendices.
 
\section{Six-Folds with $SU(3)$ Structure}

In this section, we give a brief review of six-folds with $SU(3)$ structure,
relevant to, e.g., Heterotic theory compactifications in the presence of
$NS-NS$ flux, and a short review of the ``half-flat"  Iwasawa
manifold.

\subsection{Brief review of torsion classes}

As per the work of \cite{strominger}, the requirements for ${\cal N}=1$ 
supersymmetric compactification of heterotic string theory are:

\noindent (a) 
The internal 6-dim manifold has to be complex. 
That means that the Nijenhuis tensor $N_{mnp}$ has to vanish.

\noindent (b) Up to a constant factor, there 
is exactly one holomorphic (3,0)-form $ \omega$ whose norm is related 
to the complex structure $J$ 
by $ \star d \star J = i (\bar{\partial}-\partial) \log{|| \omega ||}$.

\noindent (c) The Yang Mills background field strength must be a (1,1)-form and 
must satisfy $tr F \wedge F =  tr \tilde{R} 
\wedge \tilde{R}-i\partial\bar{\partial}J$ as well as $ F_{mn}J^{mn}=0$ (the
Donaldson-Uhlenbeck-Yau condition), $\tilde{R}$ being the modified 
curvature two-form
in the presence of torsion.

\noindent (d)
The Warp factor is given by $ {\bigtriangleup}(y) = \phi(y) + {\rm const}$ ;
the dilaton by $ \phi(y)=\frac{1}{8} \log{|| \omega ||} + {\rm const}$. 

\noindent (e) 
The   background 3-form $H$ is determined in terms of $J$ 
by $H = {\frac{i}{2}}(\bar{\partial}-\partial)J .$
 
It is possible to reformulate these conditions in terms of torsional 
constraints \cite{Cardosoetal,cs}.
  NS-NS flux $\neq0 $ requires $\tilde{Y}$ to be a manifold with $SU(3)$ 
structure but not $ SU(3) $ holonomy.
 $SU(3)$ structure implies: 
$\exists$ J, $ \Omega $ such that $ dJ \neq{0}$ ; $ d{\Omega}\neq{0}$ 
which means that the Manifold is not K\"{a}hler and then can also not be CY.
$SU(3)$-structure can now be determined in terms of torsion classes.

The difference between any two metric-compatible connections 
is a tensor, known as
the contorsion ${\kappa}_{mnp}$ defined via: 
$${\bigtriangledown}_m^{(T)} \eta=
{\bigtriangledown}_m\eta-\frac{1}{4}{\kappa}_{mnp} {\Gamma}^{np}\eta =0,$$
$\eta$ being the globally defined spinor that is covariantly constant w.r.t.
the connection modified by the the three-form flux $H$.
The contorsion can be related to torsion $T_{mnp}$ through
$$T_{mnp}=\frac{1}{2}({\kappa}_{mnp}-{\kappa}_{nmp}).$$
 The torsion classes can be defined in terms of J, $ \Omega $, dJ, $ d{\Omega}$ and
the contraction operator  $\_| : {\Lambda}^k T^{\star} \otimes {\Lambda}^n
T^{\star} \rightarrow {\Lambda}^{n-k} T^{\star}$
where $J$ is given by:
$$ J  =  e^1 \wedge e^2 + e^3 \wedge e^4 + e^5 \wedge e^6, $$
and 
the (3,0)-form $ \Omega $ is given by 
$$ \Omega  =  ( e^1 + ie^2) \wedge (e^3 +
ie^4) \wedge (e^5 + ie^6). $$
The basis of $(1,0)$-forms is given by
$${e^i+iJ \cdot e^i}\in{\Lambda}^{(1,0)},$$
where $J \cdot e^a ={J^a}_be^b$ and consequently $J \cdot J=-1$.
 
The torsion classes are defined in the following way:
$W_1 \leftrightarrow [dJ]^{(3,0)}$, given by real numbers 
$W_1=W_1^+ + W_1^-$
with $ d {\Omega}_+ \wedge J = {\Omega}_+ \wedge dJ = W_1^+ J\wedge J\wedge J$
and $ d {\Omega}_- \wedge J = {\Omega}_- \wedge dJ = W_1^- J \wedge J \wedge J$;
similarly for $W_2 \leftrightarrow [d \Omega]_0^{(2,2)}$ : 
$(d{\Omega}_+)^{(2,2)}=W_1^+ J \wedge J + W_2^+ \wedge J$
and $(d{\Omega}_-)^{(2,2)}=W_1^- J \wedge J + W_2^- \wedge J$;
 $W_3 \leftrightarrow [dJ]_0^{(2,1)}$ is defined 
as $W_3=dJ^{(2,1)} -[J \wedge W_4]^{(2,1)}$;
 $W_4 \leftrightarrow J \wedge dJ$ : $W_4 =\frac{1}{2} J\_|dJ$;
 $W_5 \leftrightarrow 
[d \Omega]_0^{(3,1)}$: $W_5 = \frac{1}{2} {\Omega}_+\_|d{\Omega}_+$
(the subscript 0 indicative of the primitivity of the respective forms).
Depending on the classes of torsion one can obtain different types of manifolds:
\begin{itemize}
\item
(complex) special-hermitian  manifolds with $ W_1=W_2=W_4=W_5=0$ which 
means that $ \tau \in W_3 $; 

\item
(complex) K\"{a}hler  manifolds with $ W_1=W_2=W_3=W_4=0$ which means 
$\tau\in W_5 $;

\item
(complex) balanced Manifolds with $W_1=W_2=W_4=0$ which 
means $\tau\in W_3\oplus W_5$;

\item
(complex) 
Calabi-Yau manifolds with $ W_1=W_2=W_3=W_4=W_5=0$ which means $\tau =0$. 
 
\item
half-flat manifolds (may or may not be complex)
 with $W_1=W_2=0$ whch means  $\tau \in W_3 \oplus W_4 \oplus W_5$.
\end{itemize}

\subsection{Heterotic string on the Iwasawa manifold}

 The torsional constraints are: 
$\tau \in W_3 {\oplus} W_4 {\oplus} W_5$;\ $2W_4+W_5=0$ and $W_{4,5}$ real, 
exact. We consider now manifolds fullfilling the torsional constraints 
and satisfying supersymmetry requirements.
From this we obtain, that we need special-hermitian manifolds for 
which the torsion is $ {\tau \in W_3} $; they are complex and half-flat.
Since $ W_4=W_5=0 $ it follows that the dilaton is constant.
We consider nilmanifolds (6-dim) which are special-hermitian manifolds which are
constructed from simply-connected nilpotent Lie group G by quotienting with 
discrete subgroup $ \Gamma $ of G for which $ G {\setminus}{\Gamma}$ is compact.

There are 34 classes of such manifold and they do not admit a K\"{a}hler metric.
18 of these admit complex structure. An example is the Iwasawa manifold.
For any of 18 classes one can choose complex structure, compatible with metric 
and $W_4=0$. Iwasawa manifold is a nilmanifold obtained as the compact quotient 
space $M=\Gamma\setminus G$,
where $G$ the complex Heisenberg group is given by a set of 
matrices under multiplication
$$ G=\Biggl\{\left(\begin{array}{*{3}{c}}
1 & z & u\\
0 & 1 & v \\
0 & 0 & 1 \\
\end{array}
\right) : u,v,z \in C \Biggr\}.$$\\
The discrete subgroup $\Gamma$ is defined by restricting $u,v,z$ to Gaussian
integers:\\
\begin{eqnarray*}
 v &\rightarrow& v+m \nonumber\\
   z &\rightarrow& z+m \nonumber\\
   u &\rightarrow& u+p+nv 
\end{eqnarray*} 
where $m,n,p \in Z \oplus iZ$.
 To find an Iwasawa manifold 
solution of torsional constraints the parameters have
to be choosen such that the torsion lies in $ \tau \in W_3 $ with $ 2W_4 + W_5 = 0$
and $ W_4= W_5=0 $.
 It turns out that the moduli space of complex structures has two disconnected
components referred as the standard complex structure $J_0$
and 'edge'.
It can be shown that in both components $W_4=W_5 =0$.

The Standard complex stucture is given by: 
$ J_0 = e^1 \wedge e^2 + e^3 \wedge e^4 + e^5 \wedge e^6 $
with the (1,0)-forms are given by
$$ \alpha = e^1 + ie^2,\ \ \beta = e^3 + ie^4,\ \ \gamma = e^5 - ie^6. $$
Complex coordinates $(z, v, u)$ can be introduced 
\begin{eqnarray*}
\alpha &=& dz,\nonumber\\
\beta &=& dv,\nonumber\\
\gamma &=& i(du - zdv).
\end {eqnarray*}
These are holomorphic left-invariant (1,0)-forms 
with respect to the standard complex structure.

 The two-form in the standard complex structure limit, is given by
$$ J_0 = \frac{i}{2}[ dz \wedge d\bar{z}
 + dv \wedge d\hat{v} + (du - zdv) \wedge (d\hat{u} - \bar{z}d\bar{v})],$$
thereby implying that the metric is 
$ds^2 = |dz|^2 + |dv|^2 + |du - zdv|^2$.
The Iwasawa manifold can thus be viewed as 
$ T^2 $ fibration over a $ T^2 \times T^2 $ base. The Euler characteristic
for the same is hence zero.
The 3-form is given by 
$\Omega =\alpha \wedge \beta \wedge \gamma = idz \wedge dv \wedge du.$

\section{Uplift}

 ${\cal N}=1$ theories in four dimensions starting from M-theory
 in the presence of G ($\equiv$ 4-form) fluxes require 7-folds with
 either $G_2$ holonomy or $SU(3)$ structure. We first discuss the uplift
of the Iwasawa to seven-folds with $G_2$ holonomy.

\subsection{ Hitchin's construction of 7-folds with $G_2$ Holonomy from 6-folds that
are half-flat}
 
From a six dimensional manifold $M$ with $SU(3)$ structure ($M,J, \Omega $) 
one can construct a 7-dim manifold as a warped product 
$$ X_7 = {M} \times I; \ {I}\subset{\bf R}. $$
A $ G_2 $ -Manifold is defined by: 
$ \phi = J \wedge dt + {\Omega}_+ $ where the calibration $\phi$ is closed 
and coclosed and $ (J,{\Omega}_+,{\Omega}_-) $ are $t$-dependent. This implies:
\begin{eqnarray}
\label{eq:G2conds}
& & d{\phi} = \biggl(\ {\hat{d}J}- \frac{\partial {{\Omega}_+}}
{\partial t}\biggr)\ \wedge dt + \hat{d}{{\Omega}_+}=0;\nonumber\\
& & d \star {\phi} =\biggl(\  \hat{d}{{\Omega}_-} - 
J  \wedge \frac{\partial J}{\partial t}\biggr)\ \wedge dt 
+ J \wedge \hat{d}J=0.
\end{eqnarray}
Through this the forms have now been promoted to seven dimensions.
 $M$ is half-flat which means that its $SU(3)$ structure is such that: 
\begin{equation}
 \hat{d}{{\Omega}_+}=J \wedge \hat{d}J =0 .
\end{equation}
The conditions (\ref{eq:G2conds}) for the seven-manifold $ X_7 $ to have
$ G_2 $ holonomy , i.e. $ d {\phi} 
= d \star {\phi} = 0 $, yield:
\begin{eqnarray}
  \hat{d}J & = & \frac{\partial {\Omega}_+}{\partial t}, \\ \hat{d}{{\Omega}_-}& = &
- J \wedge \frac{\partial J}{\partial t}  
\end{eqnarray}
called Hitchin's Flow equations.

\subsection{Uplift of the Iwasawa manifold to $G_2$-holonomy manifold 
via Hitchin's Flow equation}

In this subsection, we discuss the uplift of the Iwasawa to a seven-fold of
$G_2$-holonomy using ``size"(implying $t$-dependent) deformation of the
Iwasawa metric. We later show that ``size" alongwith nontrivial ``shape"
(implying $z,{\bar z}; v, {\bar v}$-dependent) deformations of the Iwasawa
metric can not be used to uplift to a $G_2$-holonomy manifold.

We use the following ansatz for $J$ with ``size" deformation:
\begin{eqnarray}
 J &=& e^{a(t)}e^{12} + e^{b(t)}e^{34} + e^{c(t)}e^{56} \\
\Rightarrow \Omega &=& e^{\frac{1}{2}(a(t)+b(t)+c(t))}(e^1+ie^2) \wedge (e^3+ie^4)
\wedge (e^5+ie^6)  
\end{eqnarray}   
 We now use the following two-step algebra:
\begin{eqnarray*}
 \hat{d}e^1=\hat{d}e^2=\hat{d}e^3=\hat{d}e^4=0;\ \ \hat{d}e^5= e^{14}+e^{23};\ 
\ \hat{d}e^6= e^{13}+e^{42}; 
\end{eqnarray*}
with the representations:
\begin{eqnarray}
& & e^1 + ie^2 = dz;\ \ e^3 + ie^4 = dv;\ \ e^5 - ie^6 = i(du-zdv).
\nonumber\\
\Rightarrow \hat{d}J &=& e^{c(t)}(e^{146}+ e^{236}-e^{513}-e^{542})  
\end{eqnarray}
 The condition $\hat{d}{\Omega}_+ = J \wedge \hat{d}J = 0 $ is identically
fullfilled.
 We solve $\hat{d}{{\Omega}_-} =  - J \wedge \frac{\partial J}{\partial t}$:
\begin{eqnarray}
& & \Rightarrow \dot{a}=\dot{b} = -\dot{c}; \nonumber\\ 
& & {d\over dt}(e^{a+b})=-4e^{a+b+c\over2}
\end{eqnarray}
These are consistent with $\hat{d}J=\frac{\partial {\Omega}_+}{\partial t}$.
 For the metric we then obtain a one-parameter ($\lambda$) family
of singular $G_2$-metrics:
\begin{equation}
 d{s_7}^2 = dt^2+ (1+\lambda t)^{\frac{2}{3}}|dz|^2 
+ (1+\lambda t)^{\frac{2}{3}}|dv|^2 + (1+\lambda t)^{-\frac{2}{3}}|du - zdv|^2.
\end{equation}
By specifying suitable boundary conditions at the end-points of the interval,
one could perhaps use such manifolds in (heterotic-)$M$-theory (like)
compactifications.

It is not possible to uplift the Iwasawa to $G_2$-manifold via both 
``size" and
(non-trivial) ``shape" deformations of the Iwasawa metric. 
We show the same now. Using e.g. the following ansatz:
\begin{eqnarray}
& &  J=e^{a(t)}e^{12} + A(z,\bar{z}, v, \bar{v})e^{b(t)}e^{34} 
+ B(z, \bar{z}, v, \bar{v})e^{c(t)}e^{56} \nonumber\\
& & \Omega=e^{\frac{a+b+c}{2}}(e^{A_1}e^1+ie^{A_2}e^2) 
\wedge (e^{A_3}e^3+ie^{A_4}e^4) \wedge(e^{A_5}e^5+ie^{A_6}e^6), 
\end{eqnarray}
with "shape" and "size" deformation we will obtain from the 
Hitchin's Flow equations, the conditions for halfflatness and 
the additional condition 
$\Sigma{A_i}=A+B$.
From the halfflatness condition $\hat{d}{\Omega}_+=0$ we obtain:
\begin{eqnarray}
& & {A_{135}}_z={A_{135}}_{\bar{z}}\nonumber\\
& & {A_{135}}_v={A_{135}}_{\bar{v}}\nonumber\\ 
& & {A_{146}}_z={A_{146}}_{\bar{z}}\nonumber\\ 
& & {A_{146}}_v=-{A_{146}}_{\bar{v}}\nonumber\\ 
& & {A_{236}}_z=-{A_{236}}_{\bar{z}}\nonumber\\ 
& & {A_{236}}_v={A_{236}}_{\bar{v}}\nonumber\\ 
& & {A_{245}}_z=-{A_{245}}_{\bar{z}}\nonumber\\ 
& & {A_{245}}_v=-{A_{245}}_{\bar{v}}, 
\end{eqnarray}
where ${A_{ijk}}_{z\ {\rm or}\ {\bar z}}\equiv{\partial(A_i+a_j+A_l)\over
\partial z({\rm or}\ {\bar z})}$.
From the halfflatness condition $J \wedge \hat{d}J=0$ we obtain:
\begin{eqnarray}
& & A_z=B_z\nonumber\\
& & A_{\bar{z}}=B_{\bar{z}}.
\end{eqnarray}
Considering the Hitchin's Flow equation 
$\frac{\partial{{\Omega}_+}}{\partial{t}}=\hat{d}J$ it follows:
\begin {eqnarray}
& & -\frac{\dot{a}+\dot{b}+\dot{c}}{2}e^{\frac{a+b+c}{2}}e^{A_{ijk}}=e^ce^B
\nonumber\\
& & (ijk={135;146;236;245})\nonumber\\
& & \Rightarrow A_{135}=A_{146}=A_{236}=A_{245}.
\end{eqnarray}
The above equations give us
\begin{equation}
 A_{135}=A_{146}=A_{236}=A_{245}= {\rm const}\label{drei}\\
\end{equation}
From the other Hitchin's Flow equation 
$\hat{{\Omega}_-}=-J \wedge \frac{\partial{J}}{\partial{t}}$ we obtain:
\begin{eqnarray}
& & \dot{a}=\dot{b}=-\dot{c}\nonumber\\
& & -e^Ae^{a+b}(\dot{a}+\dot{b})=e^{A_{136}}+e^{A_{145}}
+e^{A_{235}}+e^{A_{246}}\nonumber\\
& & {A_{136}}_z={A_{136}}_{\bar{z}}\nonumber\\
& & {A_{136}}_v={A_{136}}_{\bar{v}}\nonumber\\ 
& & {A_{145}}_z={A_{145}}_{\bar{z}}\nonumber\\ 
& & {A_{145}}_v=-{A_{145}}_{\bar{v}}\nonumber\\ 
& & {A_{235}}_z=-{A_{235}}_{\bar{z}}\nonumber\\ 
& & {A_{235}}_v={A_{235}}_{\bar{v}}\nonumber\\ 
& & {A_{246}}_z=-{A_{246}}_{\bar{z}}\nonumber\\ 
& & {A_{246}}_v=-{A_{246}}_{\bar{v}}.
\end{eqnarray}
We can now use the additional condition:
\begin{equation}
 A_{135}+A_{146}+A_{236}+A_{245}=2\Sigma{A_i}=2(A+B)={\rm const}
\end{equation}
Thus $A$ has to be a constant what tells us, that there cannot be a 
``shape" deformation in $J$.
Now we can use $\Sigma{A_i}={\rm const}$ again and write:
\begin{eqnarray}
& & {A_{123456}}_{z;\bar{z};v;\bar{v}}=(A_{136} +
A_{245})_{z;\bar{z};v;\bar{v}}=(A_{145} + A_{236})_{z;\bar{z};v;\bar{v}} \nonumber\\
& &=(A_{235} + 
A_{146})_{z;\bar{z};v;\bar{v}}=(A_{246} + A_{135})_{z;\bar{z};v;\bar{v}}=0.
\end{eqnarray}
Using equation (\ref{drei}), it follows:
\begin{equation}
A_{136}=A_{145}=A_{235}=A_{246}={\rm const}
\end {equation}
That means that we also cannot have ``shape" deformation in $\Omega$. 
The uplift of Iwasawa manifold to $G_2$ holonomy can only be done with 
``size" deformation.

\subsection{Intrinsic torsion classes for M-theory with fluxes 
in 7-dimension with $SU(3)$ structure}

The difference compared to 6-dimensional case is the existence of a 
globally defined
vector $v$. $SU(3)$ structure in d=7 is then described by a 
triplet $v, J, \Omega$.
The 2-form $J$ now satisfies $ J_a^bJ_b^c = -{\delta}_a^c + v_av^c$.
The metric of seven-dim space can be written as 
$ d{s_7}^2(x,t) = d{s_6}^2(x,t) + v \otimes v$ with $v = e^{q{\phi}(x)}dt$.
 In 7 dimensions the decomposition of torsion gives us 15  classes:
$$ \tau \rightarrow R \oplus C_{1,2} \oplus V_{1,2} \oplus W_{1,2} 
\oplus S_{1,2} \oplus {\cal A}_{1,2} \oplus T $$
where $C_{1,2}, W_{1,2}$ and T are complex.

 After using necessary and sufficient conditions for obtaining $N=1$ 
supersymmetry solutions of M theory with fluxes, and comparing with the 
decomposition of the 4-form $G$-flux given in terms of irreducible $SU(3)$
representations:
\begin{equation}
G=-{Q\over6}J\wedge J + J\wedge {\cal A} + \Omega_-\wedge V + v\wedge(
\alpha_1\Omega_+ + \alpha_2\Omega_-+J\wedge W + U),
\end{equation}
and the corresponding expression for $*_7G$ (that involves an
additional flux component ``$S$'') we obtain:
\begin{eqnarray*}
R &=& C_1=W_1 = W_2 = {\cal A}_1 = T = S_2 =0;\nonumber\\
C_2 &=& \bar{C}_2 =0;\nonumber\\
V_1 &=& \frac{2}{3} V_3 = \sigma;\nonumber\\
{\cal A}_2 &=& -{\cal A};\nonumber\\
S_1&=&-2S.\nonumber\\
\end{eqnarray*}

\subsection{Irreducible 7-manifold}
 
A naturally induced $SU(3)$-structure on seven-dimensional 
manifold without a killing isometry is given by:
$$ v=e^{q \phi}dt, \ J=\hat{J}(t),\ \Omega = \hat{\Omega}(t)$$
where $t$ is variable parameterizing the interval $I$.
For simplicity we first discuss $q=0$:
$$ d{s_7}^2(y,t) = d{s_6}^2(y,t) + dt^2.$$

The conditions $\hat{d}J=-2S\neq0,\ \hat{d}\Omega=0$ give us that
$M_6$ is a special-hermitian manifold,i.e. it 
is a complex non-K\"{a}hler manifold \footnote{As the torsion classes
$W_1$ and $W_2$ are defined via: $(\hat{d}\hat{\Omega}_\pm)^{(2,2)}
=W_1^\pm\hat{}\wedge\hat{J}+W_2^\pm\hat{J}$, this implies $W_1=W_2=0$,
implying the vanishing of the Nijenhuis tensor. Hence the manifold is a 
complex manifold. Further,  $W_5={1\over2}\hat{\Omega_+}\_|\hat{d}
\hat{\Omega_+}=0$. As $W_1$ is also identified 
with $[\hat{d}\hat{J}]^{(3,0)}$ and
given that $W_1=0$ this implies that $dJ\in\Lambda^{(2,1)}\oplus\Lambda^{(1,2)}
$[As $\hat{d}\hat{J}\neq0$, writing $\alpha^{(p,q)}$ in terms of a real
$\alpha^{p+q}$: 
$\alpha^{(p,q)}_{m_1.....m_{p+q}}=(P^+)^{n_1}_{m_1}...(P^+)^{n_p}_{m_p}
(P^-)^{n_{p+1}}_{m_{p+1}}....(P^-)^{n_{p+q}}_{m_{p+q}}
\alpha^{p+q}_{n_1...n_{p+q}}$, where $(P^\pm)^n_m\equiv{1\over2}(\delta^n_m
\pm iJ^n_m)$, this implies that $\hat{d}\alpha^{(p,q)}\in
\Lambda^{(p+2,q-1)}\oplus\Lambda^{(p+1,q)}\oplus\Lambda^{(p,q+1)}\oplus
\Lambda^{(p-1,q+2)}$\cite{gurrierietal}, where $\alpha^{(p,q)}\in\Lambda^{(p,q)}$.]. Further,
as $W_3$ is identified with $[\hat{d}\hat{J}]^{(2,1)}$ and 
$\hat{d}\hat{J}^{(2,1)}=(\hat{J}\wedge W_4)^{(2,1)}+W_3$, one concludes
that $W_4=0$. Thus, the six-fold is a special hermitian manifold.}.
 One can now build a 7-manifold, that can be used in flux solutions of 
M-theory by solving dif\~ferential equations in $t$.
This construction generalizes the Hitchin's Flow equations for
construction of $G_2$-holonomy man\~ifold:
\begin{eqnarray}
& & \frac{\partial J}{\partial t}= \frac{2}{3} Q J -2{\cal A}\nonumber\\
& & \frac{\partial \Omega}{\partial t}= Q \Omega.
\end{eqnarray}

We  now use an ansatz for $J$ with ``shape" and ``size" deformations:
\begin{equation}
\label{eq:ansatzJ}
 J=e^{a(t)}e^{12} + A(z,\bar{z}, v, \bar{v})e^{34} + B(z, \bar{z}, v, \bar{v})e^{56} 
\end{equation}
where $Im(A)= Im(B)=0$.
\begin{eqnarray}
  \Rightarrow S &=&-\frac{1}{2}\hat{d}J 
= -\frac{1}{2}[(e^{134} + ie^{234}) \frac{\partial A}{\partial z} 
+ (e^{134} - ie^{234}) \frac{\partial A}{\partial \bar{z}}\nonumber\\
& &  + (e^{156} + ie^{256}) \frac{\partial B}{\partial z} 
+ (e^{156} - ie^{256}) \frac{\partial B}{\partial \bar{z}}
+ (e^{356} + ie^{456}) \frac{\partial B}{\partial v} \nonumber\\
& & + (e^{356} - ie^{456}) \frac{\partial B}{\partial \bar{v}}
+ B(-e^{135} + e^{245} + e^{146} + e^{236})] 
\end{eqnarray}

The Bianchi identity reads:
\begin{eqnarray}
& & 
\frac{dQ}{6}\wedge J \wedge J + \frac{2}{9} Q^2 J\wedge J\wedge v  - \frac{4}{3} Q
\,v\wedge J\wedge {\cal A} -
\frac{Q}{3} J\wedge J \wedge \sigma \nonumber\\
& &  +2 \,S\wedge {\cal A} + 2\, v\wedge
{\cal A}\wedge {\cal A} 
+ J \wedge \sigma\wedge  {\cal A} - 3\, v\wedge \sigma\wedge
J\wedge
W\nonumber\\
& &  - 2\, v\wedge S\wedge W + v\wedge J\wedge dW + d*S \stackrel{?}{=} 0\,
\end{eqnarray}
where the question mark over the equality sign is indicative of the possibility
that one can not satisfy a source-free Bianchi identity (as will be the
case).

With $QJ= \frac{3}{2} [\dot{J} + 2A] \Rightarrow Q=\frac{3}{2} \dot{a}(t)$ it
follows:
\begin{eqnarray}
& &  \frac{dQ}{6} \wedge J \wedge J = \frac {\ddot{a}}{2} dt 
\wedge[e^a A e^{1234} + e^a B e^{1256} +  A B e^{3456} ] \\
& &  \frac{2}{9} Q^2 J \wedge J \wedge v= {\dot{a}}^2 dt \wedge[e^a A e^{1234} + e^a
B e^{1256} +  A B e^{3456} ]
\end{eqnarray} 
And with $ {\cal A} \equiv \frac{1}{2} \dot{a}(t) [A(z,\bar{z}, v, \bar{v})e^{34} + 
B(z, \bar{z}, v, \bar{v})e^{56} ]$ it follows:
\begin{eqnarray}
& &  -\frac{4}{3} Q v \wedge J \wedge A= -{\dot{a}}^2 dt \wedge[e^a A e^{1234} + e^a
B e^{1256} + 2 A B e^{3456} ]\\
\nonumber\\
& &  2 S \wedge {\cal A}  = 
-\frac{\dot{a}(t)}{2}[e^{13456}(\frac{\partial}{\partial z} +
\frac{\partial}{\partial \bar{z}}) (AB) + ie^{23456}(\frac{\partial}{\partial z} -
\frac{\partial}{\partial \bar{z}}) (AB)]\\
\nonumber\\
& &  2 v \wedge {\cal A} \wedge {\cal A}= {\dot{a}}^2 AB dt \wedge e^{3456}
\end{eqnarray}
\begin{eqnarray}
& &   d{\star}_7 S =
 -\frac{1}{2} dt \wedge [2e^{1256}(\frac{{\partial}^2 A}
{{\partial z}^2} + \frac{{\partial}^2 A}{{\partial \bar{z}}^2})\nonumber\\
& & +2e^{1234}(\frac{{\partial}^2 B}{{\partial z}^2} + \frac{{\partial}^2
B}{{\partial \bar{z}}^2} +\frac{{\partial}^2 B}{{\partial v}^2} + \frac{{\partial}^2
B}{{\partial \bar{v}}^2}+2B)\nonumber\\
& & -e^{2356}(\frac{{\partial}^2 A}{{\partial z}{\partial v}} + \frac{{\partial}^2
A}{{\partial \bar{z}}{\partial v}} + c. c.)
 -ie^{2456}(\frac{{\partial}^2 A}{{\partial z}{\partial v}} + \frac{{\partial}^2
A}{{\partial \bar{z}}{\partial v}} - c. c.)\nonumber\\
& & -ie^{1356}(\frac{{\partial}^2 A}{{\partial z}{\partial v}} - \frac{{\partial}^2
A}{{\partial \bar{z}}{\partial v}} - c. c.)
+ e^{1456}(\frac{{\partial}^2 A}{{\partial z}{\partial v}} - \frac{{\partial}^2
A}{{\partial \bar{z}}{\partial v}} + c. c.)\nonumber\\
& & i(e^{1346} + e^{2345})(\frac{\partial B}{\partial v}-\frac{\partial B}{\partial
\bar{v}}) -(e^{1345} - e^{2346})(\frac{\partial B}{\partial v}+\frac{\partial
B}{\partial \bar{v}})\nonumber\\
& & -i (e^{1245} + e^{1236})[(\frac{\partial B}{\partial z}-\frac{\partial
B}{\partial \bar{z}})+(\frac{\partial A}{\partial z}-\frac{\partial A}{\partial
\bar{z}})]\nonumber\\
& & + i(-e^{1246} + e^{1235})[(\frac{\partial B}{\partial z}+\frac{\partial
B}{\partial \bar{z}})-(\frac{\partial A}{\partial z}+\frac{\partial A}{\partial
\bar{z}})]]
\end{eqnarray}
The Hodge dual of $dG$ then reads:
\begin{eqnarray}
& & {{\star}_{11}}dG 
= e^{4\Delta}dx^0\wedge dx^1\wedge dx^2\wedge
dx^3 \wedge \nonumber\\ 
& & \biggl[(\frac{\ddot{a}}{2} e^aA -(\frac{{\partial}^2 B}{{\partial z}^2} +
\frac{{\partial}^2 B}{{\partial \bar{z}}^2} +\frac{{\partial}^2 B}{{\partial v}^2} +
\frac{{\partial}^2 B}{{\partial \bar{v}}^2}+2B)) e^{56}\nonumber\\
&+& (\frac{\ddot{a}}{2} e^aB -(\frac{{\partial}^2 A}
{{\partial z}^2} + \frac{{\partial}^2 A}{{\partial \bar{z}}^2})) e^{34}
+ \frac{\ddot{a}}{2}AB e^{12}- 
\frac{\dot{a}(t)}{2}(\frac{\partial}{\partial z} + \frac{\partial}{\partial
\bar{z}}) AB dt \wedge  e^2 \nonumber\\ 
&& 
-i\frac{\dot{a}(t)}{2}(\frac{\partial}{\partial z} - \frac{\partial}{\partial
\bar{z}}) AB dt \wedge  e^1 \nonumber\\ 
&& -\frac{1}{2}  [ -e^{14}(\frac{{\partial}^2 A}{{\partial z}{\partial v}} +
\frac{{\partial}^2 A}{{\partial \bar{z}}{\partial v}} + c. c.)
 -ie^{13}(\frac{{\partial}^2 A}{{\partial z}{\partial v}} + \frac{{\partial}^2
A}{{\partial \bar{z}}{\partial v}} - c. c.)\nonumber\\
& & -ie^{24}(\frac{{\partial}^2 A}{{\partial z}{\partial v}} - \frac{{\partial}^2
A}{{\partial \bar{z}}{\partial v}} - c. c.)
+ e^{23}(\frac{{\partial}^2 A}{{\partial z}{\partial v}} - \frac{{\partial}^2
A}{{\partial \bar{z}}{\partial v}} + c. c.)\nonumber\\
& & i(e^{25} + e^{16})(\frac{\partial B}{\partial v}-\frac{\partial B}{\partial
\bar{v}}) -(e^{26} - e^{15})(\frac{\partial B}{\partial v}+\frac{\partial
B}{\partial \bar{v}})\nonumber\\
& & -i (e^{36} + e^{45})[(\frac{\partial B}{\partial z}-\frac{\partial B}{\partial
\bar{z}})+(\frac{\partial A}{\partial z}-\frac{\partial A}{\partial
\bar{z}})]\nonumber\\
& & + i(-e^{35} + e^{46})[(\frac{\partial B}{\partial z}+\frac{\partial B}{\partial
\bar{z}})-(\frac{\partial A}{\partial z}+\frac{\partial A}{\partial
\bar{z}})]]\biggr]
\end{eqnarray}
We use this form because $dG = {\star}_{11} {\cal I}_6$, the six-form 
${\cal I}_6$ specifying the position of the $M5$-branes transverse to the world
volume, when one interprets the metric to represent $M5$-branes wrapped
around two-cycles (with densities $\rho_{ij}$) (See \cite{dp}). 
Computing ${{\star}_{11}}dG$ we obtain informations about ${\cal I}_6$. 
 
We now simplify in the following way: 
$A, B$ do not depend on $v$ and $\bar{v}$; $A=B$;
$\frac{\partial A}{\partial z}= \frac{\partial A}{\partial \bar{z}}$.
One parameter family of solution satisfying the assumptions and the 
requirement of periodicity in the $T^2$-valued $z$, is given by
 $A(z,{\bar z})= cos[m\pi (z + \bar{z})]$ 
where $m\in{\bf Z}$.

The Hodge dual of $dG$ then reads:
\begin{eqnarray}
& & {{\star}_{11}}dG = e^{4\Delta}dx^0\wedge dx^1\wedge dx^2\wedge
dx^3 \wedge \nonumber\\
&& \biggl[\biggl(\frac{\ddot{a}}{2} e^a + 2(m\pi)^2-2\biggr) Ae^{56}
+ \biggl(\frac{\ddot{a}}{2} e^a + 2(m\pi)^2\biggr) Ae^{34}
 +\frac{\ddot{a}}{2}A^2 e^{12} 
- 2 \alpha \dot{a} {\partial A\over\partial z}  e^{2t}\biggr]\nonumber\\ 
& & 
\equiv e^{4\Delta}dx^0\wedge dx^1\wedge dx^2\wedge
dx^3 \wedge
({\rho}_{12} e^{12} + {\rho}_{34} e^{34} + {\rho}_{56} e^{56} + {\rho}_{2t} e^{2t}) 
\end{eqnarray}
{\it Interestingly, along hypersurfaces given by: $z+{\bar z}={(2k+1)\over 2m}$,
one gets a source-free Bianchi identity}. For other points,
we consider now different cases for ${{\star}_{11}}dG$:

\begin{itemize}
\item
 $a={\rm const}$, which would mean that we do not have "size" deformation.
$\Rightarrow {{\star}_{11}}dG =  - 2 {\alpha}^2 Ae^{34}
+2(m\pi)^2-2) Ae^{56}$ 
That means the densities ${\rho}_{34}$ and ${\rho}_{56}$ are not zero; 
it is the only posibility that two terms vanish.

\item
$\dot{a}(t)={\rm const}$: $\rho_{12}=0$

\item
We can also require that $\rho_{34}=0$ -  we then have non-zero density 
 ${\rho}_{12}$ with $a(t)$ given by:
$a(t)=ln[{e^{-\gamma_1 t + \gamma_2}(-32(m\pi)^2 + e^{\gamma_1 t + \gamma_2})^2
\over 16\gamma_1^2}]$, $\gamma_{1,2}$ being constants of integration.
One can get a similar expression for $a(t)$ if one sets $\rho_{56}=0$.
\end{itemize}

The Warp-factor can now be calculated to:
$$\bigtriangleup(t) = - \frac{1}{2} a(t) + {\rm const}; 
\ \mbox{set}\ {\rm const}=0$$
 For the metric it follows:
$$d{s_{11}}^2=e^{-a(t)}{\eta}_{\mu \nu} dx^{\mu} dx^{\nu} + d{s_7}^2.$$

\subsection{Discussion for ${q}\neq{0}$}

For the 7-dimensional metric we assume 
$d{s_7}^2(y,t)=e^{p\phi}d{s_6}^2(y,t)+e^{2\phi}dt^2$.
One can also show the following relations:
\begin{eqnarray}
& & {\sigma}\equiv{\hat{d}\bigtriangleup}=-\frac{1}{2} \hat{d}\phi;\ \
\dot{\bigtriangleup}= -\frac{1}{3}Qe^{\phi}.
\end{eqnarray}
Unlike what is shown in \cite{dp}, 
the following equation is identically fullfilled:
\begin{eqnarray}
& &  3d\sigma=dQ \wedge v+2Qv \wedge \sigma.\nonumber\\
\end{eqnarray}
The two- and the three-form of 
SU(3) structure on the six-fold  are now:
\begin{eqnarray}
& & J=e^{p\phi} \hat J ;\ \ 
\Omega=e^{\frac{3}{2}p\phi} \hat \Omega.
\end{eqnarray}
Thus, one gets a non-K\"{a}hler complex manifold referred to as a balanced
manifold\footnote{This time, $W_5={1\over2}\hat{\Omega_+}\_|\hat{d}
\hat{\Omega_+}={3\over8}\hat{d}\phi\neq0$. Hence, the torsion $\tau\in
W_3\oplus W_5$ - a balanced manifold.}.
We obtain:
\begin{eqnarray}
& &   \hat{d} \hat{J} = -2 e^{-\frac{1}{2} \phi}S ;\ \ 
\hat{d} \hat{\Omega} = \frac{3}{4} \hat{d} \phi \wedge \hat{\Omega}
\end{eqnarray}
and for the Flow equations it follows:
\begin{eqnarray} 
 & & \frac{\partial \hat{J}}{\partial t}= \frac{2}{3} 
e^{\phi}Q \hat{J} -2e^\frac{\phi}{2}{\cal A}-\frac{1}{2} \dot{\phi} \hat{J}
\nonumber\\
 & &\frac{\partial \hat{\Omega}}
{\partial t}= Q e^{\phi} \hat{\Omega}-\frac{3}{4} \dot{\phi} \hat{\Omega}.
\end{eqnarray}\\
For simplicity we choose $p=\frac{1}{2}$. 
We assume $A=\frac{1}{2}e^{-\phi} \dot{a}(Ae^{34}+Be^{56})$ from 
where
it follows $Q=\frac{3}{2} \dot{a} e^{-\phi}$. Hence, $\Delta=-{1\over2}a(t)
-{1\over2}\phi$.
Through comparison of the coefficients using 
$\sigma=-\frac{1}{2} \hat{d} \phi = \frac{2}{3} W\_|J$ we obtain :
\begin{eqnarray}
& & W= - \frac{3}{4} [e^1i({\phi}_z-{\phi}_{\bar{z}})e^{-a}-e^2
({\phi}_z+{\phi}_{\bar{z}})e^{-a}
+ e^3i\frac{({\phi}_v-{\phi}_{\bar{v}})}{A}-e^4 
\frac{({\phi}_v+{\phi}_{\bar{v}})}{A}].
\end{eqnarray}
Now we can calculate the terms for the Bianchi identity. The same is done
in Appendix A.
The $D=11$-Hodge dual of $dG$ can then be calculated, as done towards
the end of Appendix A. 
Now, 
using the simplifications we already used for q=0:
$A, B$ do not depend on $v$ and $\bar{v}$; 
 $A=B$; $\frac{\partial A}{\partial z}= \frac{\partial A}{\partial \bar{z}}$,
thereby obtaining a 
one-parameter family of solution satisfying the assumptions
 $A(z,{\bar z})= cos[m\pi(z + \bar{z})],\ m\in{\bf Z}$, and
additionally assuming $\phi$ does not depend on $v$ and $\bar{v}$,
the eleven-dimensional Hodge dual of $dG$ then reads:
\begin{eqnarray}
& & {{\star}_{11}}dG = e^{4\Delta}dx^0\wedge dx^1\wedge dx^2\wedge
dx^3 \wedge \nonumber\\
& & \biggl[(e^{-\phi}\frac{\ddot{a}}{2} e^a - 2 {\alpha}^2 -2
+\frac{\phi_{zz}+{\phi}_{\bar{z} \bar{z}}}{2} + {\alpha}({\phi_z}
+\phi_{\bar{z}}) + \frac{3}{2}e^{\phi}e^{-a}\alpha(\phi_z+\phi_{\bar{z}})
-{1\over2}\dot{a}\dot{\phi}e^{-\phi}) Ae^{56}\nonumber\\
&+& (e^{-\phi}\frac{\ddot{a}}{2} e^a 
- 2 {\alpha}^2+\frac{\phi_{zz}+{\phi}_{\bar{z}\bar{z}}}{2} 
+ {\alpha}({\phi_z}+\phi_{\bar{z}}) +
\frac{3}{2}e^{\phi}e^{-a}\alpha(\phi_z+\phi_{\bar{z}})-{1\over2}\dot{a}
\dot{\phi}e^{-\phi}) Ae^{34} \nonumber\\
&+& (\frac{\ddot{a}}{2} - {1\over2}\dot{a}\dot{\phi}e^{-\phi})A^2 e^{12} 
+ \frac{A}{4}(\phi_z +\phi_{\bar{z}})[3e^{\phi}e^{-a}-1](e^{35}-e^{46})
\nonumber\\
&+& \frac{A}{4}(\phi_z -\phi_{\bar{z}})[3e^{\phi}e^{-a}-1]i(e^{36}+e^{45})
- 2 \alpha \dot{a}e^{-\phi} A^2 dt \wedge  e^2\biggr]\nonumber\\
& & \equiv {\rm vol\ form}({\bf R}^{3,1})(\rho_{12}e^{12} +
\rho_{34}e^{34}+\rho_{56}e^{56}+\rho_{35}e^{35}+\rho_{36}e^{36}+\rho_{45}e^{45}
+\rho_{46}e^{46}+\rho_{2t}e^{2t}).\nonumber\\
& & 
\end{eqnarray}
Generically, one thus sees, that the deformed Iwasawa  
is a balanced manifold. However, if one assumes that
$\hat{d}\phi=0$ and $\phi=\phi(t)$, then the balanced six-fold
becomes a special-hermitian six-fold. In such a situation, 
$\rho_{35}=\rho_{36}=\rho_{45}=\rho_{46}=0$. 
Again, {\it along hypersurfaces given by: $z+{\bar z}={(2k+1)\over 2m}$,
one gets a source-free Bianchi identity}. 

For a balanced manifold, assuming $\dot{\phi}=0, \phi_z=-\phi_{\bar z}$, 
which is satisfied by, e.g.,
$\phi=ln[cos[i n\pi(z-{\bar z})]],\ n\in{\bf Z}$ (once again ensuring
periodicity w.r.t. the $T^2$-valued $z$), 
one gets the following metrics for seven-folds with $SU(3)$ structure:
\begin{itemize}
\item
having frozen the ``size" deformation of the Iwasawa, the seven-fold is
a fibration of an interval over a balanced six-fold:
\begin{equation}
ds_7^2=(cos[2n\pi{\rm Im}(z)])^2dt^2 + |dz|^2 + cos[2m\pi{\rm Re}(z)]|dv|^2
+ cos[2m\pi{\rm Re}z]|du - v dz|^2,
\end{equation}
which gives
\begin{equation}
ds_{11}^2=sec[2n\pi{\rm Im}(z)]ds^2_{{\bf R}^{3,1}}+ ds_7^2;
\end{equation}
this implies that $\rho_{12}=\rho_{t2}=0$. The interval, for this case,
could also be replaced by $S^1$ as periodicity w.r.t. this $S^1$ will then
not be a problem.

\item
For the six-folds to be special-hermitian manifolds by assuming 
$\phi=\phi(t)=a(t)$
(i.e., $\hat{d}\phi=0$), and further $\ddot{a}-(\dot{a})^2=2-2(m\pi)^2$,
solved by $a(t)=\pm\biggl(\sqrt{2(m\pi)^2-2}\biggr)t + {\rm constant}$,
implying $\rho_{56}=0$, one gets the seven-fold to be a warped product of a
balanced manifold and an interval:
\begin{eqnarray}
ds_7^2& = & e^{-2\biggl(\sqrt{2(m\pi)^2-2}\biggr)t +\gamma_3}dt^2 + 
e^{-\biggl(\sqrt{2(m\pi)^2-2}\biggr)t + {\gamma_3\over2}}|dz|^2 
+ cos[2m\pi{\rm Re}(z)]|dv|^2 \nonumber\\
& & + 
cos[2m\pi{\rm Re}(z)]|du - v dz|^2,
\end{eqnarray}
which gives a two($m\in{\bf Z},\gamma_3\in{\rm R}$))-parameter family of
solutions:
\begin{equation}
ds_{11}^2=
e^{\biggl(\sqrt{2(m\pi)^2-2}\biggr)t - 
{\gamma_3\over2}}ds^2_{{\bf R}^{3,1}} + ds_7^2.
\end{equation}

\end{itemize}

\section{Interpretation of the Uplifts}

For uplifts to seven-folds with $SU(3)$ structure, 
as $dG$, generically, is not zero, we only have solutions with sources.
The non-zero piece of $dG$, for $q=0$, 
in the ansatz we worked with, could be interpreted due to 
$M5$-branes wrapping 2-cycles with respect to the 
standard complex structure in the internal seven-fold 
viewed as a warped product of a 
(special hermitian or balanced) six-fold and the internal. 
For ${q}\neq{0}$, in general, one ends up with balanced manifolds. However,
the issue of supersymmetry of the wrapped M5-branes should be looked into
more carefully, by , e.g., looking at the world-volume theory of these
branes\footnote{AM thanks O.Lunin for a discussion on this point.} to ensure
that the wrapped $M5$-branes that minimize the energy functional, belong
to $M$-theory.

The two-cycle is calibrated w.r.t.
the generalized calibration \cite{gutowskietal,gauntlettetal} $J$ 
thereby guaranteeing the minimization of the corresponding energy functional, 
but {\it away} from the 
standard-complex-structure point in the moduli space of almost complex 
structures on the Iwasawa. At any generic point in the moduli space of
almost complex structures on the Iwasawa manifold, the two form $J$ can be
described by using a basis constructed of $SO(4)$ matrices 
$P^i$ and the one forms
$e_i$ of the orthonormal basis \cite{Cardosoetal}: 
$$J=\frac{i}{2}(\alpha \wedge \bar{\alpha}+ \beta \wedge \bar{\beta}
+\gamma \wedge\bar{\gamma})$$
where:
\begin{eqnarray*} 
& &\alpha \equiv \cos{\theta}f^1-\sin{\theta}e^6+if^2\nonumber\\
& & \beta \equiv f^3 + if^4\nonumber\\
& & \gamma \equiv + i(\cos{\theta}e^6-\sin{\theta}f^1+if^2),
\end{eqnarray*}
and
$f^i={P^i}_je^j,\ P\in SO(4)$.

The wedge product of two different $f$s is defined as:
$$ f^i \wedge f^j = \frac{1}{2} {P^i}_{[k}{P^j}_{l]}e^k \wedge e^l.$$
This means that we cannot have a non-trivial "shape" (and "size")
deformation using the standard complex structure. 
By turning on fluxes one moves away from the standard complex structure.
The moduli space is then neither "edge" nore standard complex structure 
with respect
to the one forms $e_i$ of the orthonormal basis. 

If one had used a singular ansatz for $J$ implying localization w.r.t.
directions transverse to the $M5$-brane world volume, and such that 
$e^{2\Delta}$ vanished at the location of the $M5$-branes
\footnote{AM thanks J.Maldacena for emphasizing this point to him.}, then
using the results of \cite{fs}, one could interpret the
11-dimensional metric after uplifting the balanced manifold to a seven-fold
with $SU(3)$ structure, as corresponding to $M5$ brane wrapped around a 
Riemann surface in ${\bf C}^3$:
\begin{equation}
ds_{11}^2=H_1^2ds_{{\bf R}^{3,1}}^2 + g_{m{\bar n}}dz^md{\bar z}^{\bar
n}
+H_2^2dt^2,
\end{equation}
one gets the required interpretation by setting $H_1=\Delta$ and
$H_2=\phi$ (See \cite{dp}) and regarding $z,v,u$ as the complex
coordinates on ${\bf C}^3$.

Now, as an interesting mathematical curiosity, 
using the results of \cite{confimrs}, we will show that one can associate
a Riemann surface to a choice of the ``shape" deformation functions. 
Consider the system of partial differential equations:
\begin{equation}
\label{eq:conf1}
{\partial A\over\partial z} = \phi B;\ {\partial B\over\partial{\bar z}}
=-\phi A,
\end{equation}
which is equivalent to:
\begin{equation}
{\partial^2 B\over\partial z\partial{\bar z}}={1\over\phi}{\partial\phi\over
\partial z}{\partial B\over\partial{\bar z}} - \phi^2B
\end{equation}
(or one could switch $B$ with $A$ and $z\leftrightarrow{\bar z}$).
We see that the following choice solves (\ref{eq:conf1}):
\begin{eqnarray}
\label{eq:conf2}
& & A_1(z,{\bar z})=sin[m\pi(z+{\bar z})],\ B_1(z,{\bar z})=
cos[m\pi(z+{\bar z})],\ \phi(z,{\bar z})=m\pi;\nonumber\\
& & A_1(z,{\bar z})=cos[m\pi(z+{\bar z})],\ B_1(z,{\bar z})=
sin[m\pi(z+{\bar z})],\ \phi(z,{\bar z})=-m\pi.
\end{eqnarray}
One can construct an ${\bf R}^3$ with coordinates::
\begin{eqnarray}
\label{eq:R31}
& & X^1 + i X^2 
= i\int_\Gamma(A_1^2 dz^\prime - B_1^2 d{\bar z}^\prime),
\nonumber\\
& & X^1 - i X^2 = i\int_\Gamma(B_1^2dz^\prime - A_1^2d{\bar z}^\prime),
\nonumber\\
& & X^3 =  -\int_\Gamma(A_1B_1dz^\prime + A_1B_1d{\bar z}^\prime),
\end{eqnarray}
and for the contour $\Gamma={(x,y)|x=y, 0<x<1}$,
\begin{eqnarray}
\label{eq:R^32}
& & X^1 + i X^2 = -x - {i\over 4m\pi} sin(4m\pi x),\nonumber\\
& & X^3 = {cos(4m\pi x)-1\over2m\pi},
\end{eqnarray}
which would be the Weierstra$\ss$ representation for the conformal
immersion in ${\bf R}^3$ with the condition of conformal immersion
(implying $g_{zz}=g_{{\bar z}{\bar z}}=0$):
\begin{equation}
\label{eq:conf3}
\biggl({\partial X^1\over\partial z}\biggr)^2
+ \biggl({\partial X^2\over\partial z}\biggr)^2
+ \biggl({\partial X^3\over\partial z}\biggr)^2 = 0,
\end{equation}
a quadric in ${\bf CP}^2$. The induced metric on the Riemann surface
is given by:
\begin{equation}
ds^2=(A_1^2 + B_1^2)^2|dz|^2=|dz|^2,
\end{equation}
which is just a $T^2$. In appendix B, we discuss a Weierstra$\ss$ representation
of conformal immersion in ${\bf C}^3$ for the linear dilaton ansatz but
for solutions that are valid only locally because of lack of periodicity
w.r.t. the $T^2$-valued coordinates.

\section{Conclusion}

We uplifted the Iwasawa to a $G_2$-manifold via Hitchin's Flow equations
using "size"-deformations, and found that an uplift via Hitchin's flow
equations (to a seven-fold with $G_2$ holonomy) with non-trivial 
"shape" deformation is not possible.
Furthermore we uplifted the Iwasawa to an irreducible 7 manifold with
$SU(3)$-structure 
through deforming the metric using "shape" and "size"-deformations via 
generalization of the Hitchin's Flow equations. 
Without simplification the seven-fold of $SU(3)$ structure turns out
to be a warped product of either a special hermtian or a 
balanced six-fold and an interval (which for cases when one freezes the
``size deformations", could also be replaced by an $S^1$).
The uplifted metric could be interpreted as 
$M5$-branes wrapping two-cycles calibrated by the generalized
calibration (like \cite{dp}), the two-form corresponding to the almost complex
structure, but {\it away} from the standard complex structure point
(in the moduli space of almost complex structures on the Iwasawa). The
supersymmetry of these wrapped membranes needs to be further looked into.
We also showed how to associate a Riemann surface (via Weierstra$\ss$ 
representation for conformal immersion in ${\bf R}^l$) 
with the ``shape deformation"
functions of seven-folds with $SU(3)$ structure and the dilaton. The 
relationship of the same to, e.g., supersymmetric two-cycles or Riemann
surfaces embedded in seven-folds of $SU(3)$ structure around which $M5$-branes
wrap, needs to be understood.

However, if
one allows for sources, as was the case, then there seems to be the
possibility of considering singular uplifts of the following type. 
Given that it is possible to impose just the requirement of (anti-)analyticity
on the (anti-)holomorphic parts of the functions 
$A(z,{\bar z}; v,{\bar v}), B(z,{\bar z}; v,{\bar v})$
and $\phi(z,{\bar z};v,{\bar v})$, remembering that the complex coordinates
$z,v$ are $T^2$-valued, one can hence introduce the doubly periodic
functions: the Weierstra$\ss$ elliptic function with double poles and
the Jacobi elliptic function with simple poles, into the metric of the
seven-fold. The elliptic surface corresponding to the Weierstra$\ss$ elliptic
function could perhaps be related to the Riemann surface relevant to the
world volume of the $M5$-brane relevant to the uplifts. See \cite{m} for
some connections relevant to this study.

\section*{Acknowledgements}

AF would like to thank the physics department of IIT Roorkee for 
the hospitality during her stay there as part of her semester-long
study leave from RWTH, Aachen, and the organizers of THEP-I (2005) at
IIT Roorkee where some preliminary results of this work were presented.
One of us (AM) would like to thank S.Chiossi, G.Papadopoulos and
J.Gauntlett for 
useful communications, and especially J.Maldacena for critical comments
on the material of the first version of this paper and O.Lunin for discussions,
the Harvard high energy theory group (and
S.Minwalla in particular) and IAS, Princeton 
for the hospitality during his stay there where 
part of this work was completed, the Department of Atomic Energy
(Board of Research in Nuclear Sciences), Govt. of India for a research
grant (under the DAE young scientist award scheme) and the organizers of
PASCOS05, Gyeongju, Korea, where some preliminary results of this work
were presented. 
We thank H.S.Solanki for participation in the initial stages of this project.

\appendix
\section{The Bianchi Identity for $q\neq0$}
\setcounter{equation}{0}
\seceqaa

In this appendix, we discuss the evaluation of the various terms in the
Bianchi identity. 
\begin{eqnarray}
 & & \frac{dQ}{6}\wedge J \wedge J 
= \frac{e^{-\phi}}{2}[\ddot{a}(e^a A e^{1234t}+ e^a B e^{1256t} + 
AB e^{3456t})\nonumber\\
& &-\dot{a}[({\phi}_z+{\phi}_{\bar{z}})ABe^{13456} 
+ i({\phi}_z-{\phi}_{\bar{z}})ABe^{23456}+ ({\phi}_v+{\phi}_{\bar{v}})e^a 
Be^{12356}\nonumber\\ 
& & +i({\phi}_v-{\phi}_{\bar{v}})e^a Be^{12456}] + \dot{\phi} [e^a A e^{1234t}
+e^a B e{1256t} + AB e^{3456t}]
\end{eqnarray}

\begin{equation}
 \frac{2}{9} Q^2 J\wedge J\wedge v
= {{\dot{a}}^2} e^{-\phi} dt 
\wedge[e^a A e^{1234} + e^a B e^{1256} +  A B e^{3456} ]
\end{equation}

\begin{equation} 
 -\frac{4}{3} Q v \wedge J \wedge A= -{\dot{a}}^2 e^{-\phi} dt \wedge[e^a A
e^{1234} + e^a B e^{1256} + 2 A B e^{3456} ]
\end{equation}

\begin{eqnarray}
& &  -\frac{Q}{3} J\wedge J \wedge \sigma
= \frac{\dot{a}}{2}e^{-\phi}[ ({\phi}_z+{\phi}_{\bar{z}})ABe^{13456} 
+ i({\phi}_z-{\phi}_{\bar{z}})ABe^{23456}\nonumber\\
& &+ ({\phi}_v+{\phi}_{\bar{v}})e^a 
Be^{12356}+ i({\phi}_v-{\phi}_{\bar{v}})e^a Be^{12456}]]
\end{eqnarray}
\begin{eqnarray}
& &  2 S \wedge {\cal A}  =-\frac{\dot{a}}{2}e^{-\phi} 
[e^{13456}[(\frac{\partial}{\partial z} 
+ \frac{\partial}{\partial \bar{z}}) AB -({\phi}_z+{\phi}_{\bar{z}})AB]\nonumber\\
& & +ie^{23456}[(\frac{\partial}{\partial z} 
- \frac{\partial}{\partial \bar{z}}) AB -({\phi}_z-{\phi}_{\bar{z}})AB] \nonumber\\
& &-e^{12356}\frac{e^a}{2}B({\phi}_v+{\phi}_{\bar{v}})-
e^{12456}i\frac{e^a}{2}B({\phi}_v-{\phi}_{\bar{v}})]
\end{eqnarray}

\begin{equation}
2 v \wedge {\cal A} \wedge {\cal A}
= e^{-\phi} {\dot{a}}^2 AB dt \wedge e^{3456}
\end{equation}

\begin{eqnarray}
& &  J \wedge \sigma\wedge  {\cal A} 
= -\frac{\dot{a}}{2}e^{-\phi}[ ({\phi}_z+{\phi}_{\bar{z}})ABe^{13456} 
+ i({\phi}_z-{\phi}_{\bar{z}})ABe^{23456}+ ({\phi}_v
+{\phi}_{\bar{v}})\frac{e^a}{2} Be^{12356}\nonumber\\
& &+ i({\phi}_v-{\phi}_{\bar{v}})\frac{e^a}{2} Be^{12456}]
\end{eqnarray}

\begin{eqnarray}
& &  - 3\ v\wedge \sigma\wedge J \wedge W
= -\frac{9}{8}e^{\phi}[-e^{1234t}[Ae^{-a}(4{\phi}_z
{\phi}_{\bar{z}}) + \frac{e^a}{A}(4{\phi}_v{\phi}_{\bar{v}})]
-e^{1256t}Be^{-a}4{\phi}_z{\phi}_{\bar{z}}\nonumber\\
& & +e^{1356t}iB[-({\phi}_z-{\phi}_{\bar{z}})({\phi}_v+{\phi}_{\bar{v}})
e^{-a} + \frac{1}{A}({\phi}_z+{\phi}_{\bar{z}})({\phi}_v-{\phi}_{\bar{v}})]
\nonumber\\
& & +e^{1456t}B[({\phi}_z-{\phi}_{\bar{z}})
({\phi}_v-{\phi}_{\bar{v}})e^{-a} 
- \frac{1}{A}({\phi}_z+{\phi}_{\bar{z}})({\phi}_v+{\phi}_{\bar{v}})]
\nonumber\\ 
& & +e^{2356t}B[({\phi}_z+{\phi}_{\bar{z}})
({\phi}_v+{\phi}_{\bar{v}})e^{-a} 
- \frac{1}{A}({\phi}_z-{\phi}_{\bar{z}})({\phi}_v-{\phi}_{\bar{v}})]
\nonumber\\
& & +e^{2456t}iB[({\phi}_z+{\phi}_{\bar{z}})({\phi}_v-{\phi}_{\bar{v}})e^{-a} - 
\frac{1}{A}({\phi}_z-{\phi}_{\bar{z}})({\phi}_v+{\phi}_{\bar{v}})]
\nonumber\\
& & -e^{3456t} \frac{B}{A} 4{\phi}_v{\phi}_{\bar{v}}]
\end{eqnarray}
\begin{eqnarray}
& &  - 2\ v\wedge S\wedge W = -\frac{3}{4}e^{\phi}dt \wedge 
[e^{1234}
[2e^{-a}[-A_z{\phi}_{\bar{z}}-A_{\bar{z}}{\phi}_z+ A{\phi}_z{\phi}_{\bar{z}}]
+\frac{e^a}{A}2{\phi}_{v}{\phi}_{\bar{v}}]
\nonumber\\
& & +e^{1256}2e^{-a}
[-B_z{\phi}_{\bar{z}}-B_{\bar{z}}{\phi}_z+ B{\phi}_z{\phi}_{\bar{z}}]
 +e^{3456}\frac{2}{A}
[-B_v{\phi}_{\bar{v}}-B_{\bar{v}}{\phi}_v+ B{\phi}_v{\phi}_{\bar{v}}]
\nonumber\\
& & +ie^{1356}
[\frac{({\phi}_v -{\phi}_{\bar{v}})}{A}[(B_z+B_{\bar{z}})-\frac{B}{2} 
({\phi}_z +{\phi}_{\bar{z}})]-e^{-a}({\phi}_z -{\phi}_{\bar{z}})
[(B_v+B_{\bar{v}})-\frac{B}{2} ({\phi}_v +{\phi}_{\bar{v}})]]
\nonumber\\
& & -e^{1456}
[\frac{({\phi}_v +{\phi}_{\bar{v}})}{A}[(B_z+B_{\bar{z}})
-\frac{B}{2} ({\phi}_z +{\phi}_{\bar{z}})]-e^{-a}({\phi}_z
-{\phi}_{\bar{z}})[(B_v-B_{\bar{v}})
-\frac{B}{2} ({\phi}_v -{\phi}_{\bar{v}})]]\nonumber\\
& & -e^{2356}[\frac{({\phi}_v -{\phi}_{\bar{v}})}{A}
[(B_z-B_{\bar{z}})-\frac{B}{2} ({\phi}_z -{\phi}_{\bar{z}})]
-e^{-a}
({\phi}_z +{\phi}_{\bar{z}})
[(B_v+B_{\bar{v}})-\frac{B}{2} ({\phi}_v +{\phi}_{\bar{v}})]]
\nonumber\\
& & -ie^{2456}[\frac{({\phi}_v +{\phi}_{\bar{v}})}{A}
[(B_z-B_{\bar{z}})-\frac{B}{2} ({\phi}_z -{\phi}_{\bar{z}})]
-e^{-a}({\phi}_z 
+{\phi}_{\bar{z}})[(B_v-B_{\bar{v}})-\frac{B}{2} 
({\phi}_v -{\phi}_{\bar{v}})]]\nonumber\\
& & +B[-ie^{-a}(e^{1236}+e^{1245})({\phi}_z
-{\phi}_{\bar{z}})-e^{-a}(e^{1246}-e^{1235})({\phi}_z +{\phi}_{\bar{z}})\nonumber\\
& &+i\frac{({\phi}_v -{\phi}_{\bar{v}})}{A}  (e^{1346}+e^{2345})+\frac{({\phi}_v
+{\phi}_{\bar{v}})}{A}(e^{2346}-e^{1345})]]
\end{eqnarray}

$dW$ can be calculated as follows:
\begin{eqnarray}
& & dW=\frac{3}{4}[ 
\dot{a}[i({\phi}_z-{\phi}_{\bar{z}})e^{-a}e^{1t}-({\phi}_z+{\phi}_{\bar{z}})e^{-a}e^{2t}]\nonumber\\
& & +[e^{1t}i(\dot{\phi_z}-\dot{\phi_{\bar z}})e^{-a} + e^{2t}(\dot{\phi_z}
+\dot{\phi_{\bar z}})e^{-a}+e^{3t}i{(\dot{\phi_v}-\dot{\phi_{\bar v}})\over A}
-e^{4t}{(\dot{\phi_v}+\dot{\phi_{\bar v}})\over A}]
-4e^{12}{\phi}_{\bar{z}z}e^{-a}]\nonumber\\
& &
-ie^{13}[({\phi}_{zv}-{\phi}_{\bar{z}v}+{\phi}_{z\bar{v}}+{\phi}_{\bar{z}\bar{v}})e^{-a}-({\phi}_{vz}-{\phi}_{\bar{v}z}+{\phi}_{v\bar{z}}-{\phi}_{\bar{v}\bar{z}})\frac{1}{A}+\frac{(A_z+A_{\bar{z}})}{A^2}({\phi}_v-{\phi}_{\bar{v}})]\nonumber\\
& &
+e^{14}[({\phi}_{zv}-{\phi}_{\bar{z}v}-{\phi}_{z\bar{v}}+{\phi}_{\bar{z}\bar{v}})e^{-a}-({\phi}_{vz}+{\phi}_{\bar{v}z}+{\phi}_{v\bar{z}}+{\phi}_{\bar{v}\bar{z}})\frac{1}{A}+\frac{(A_z+A_{\bar{z}})}{A^2}({\phi}_v+{\phi}_{\bar{v}})]\nonumber\\
& &
+e^{23}[({\phi}_{zv}+{\phi}_{\bar{z}v}+{\phi}_{z\bar{v}}+{\phi}_{\bar{z}\bar{v}})e^{-a}-({\phi}_{vz}-{\phi}_{\bar{v}z}-{\phi}_{v\bar{z}}-{\phi}_{\bar{v}\bar{z}})\frac{1}{A}+\frac{(A_z-A_{\bar{z}})}{A^2}({\phi}_v-{\phi}_{\bar{v}})]\nonumber\\
& &
+ie^{24}[({\phi}_{zv}+{\phi}_{\bar{z}v}-{\phi}_{z\bar{v}}-{\phi}_{\bar{z}\bar{v}})e^{-a}-({\phi}_{vz}+{\phi}_{\bar{v}z}-{\phi}_{v\bar{z}}-{\phi}_{\bar{v}\bar{z}})\frac{1}{A}+\frac{(A_z-A_{\bar{z}})}{A^2}({\phi}_v+{\phi}_{\bar{v}})]\nonumber\\
& &
+e^{34}[-4{\phi}_{\bar{v}v}+\frac{2A_v{\phi}_{\bar{v}}+2A_{\bar{z}}{\phi}_v}{A^2}]\end{eqnarray}

\begin{eqnarray}
& &  v\wedge J\wedge dW\nonumber\nonumber\\
& &=
\frac{3}{4}e^{\phi}dt \wedge[ B[-e^{1256}4{\phi}_{\bar{z}z}e^{-a}]\nonumber\\
& &
-ie^{1356}[({\phi}_{zv}-{\phi}_{\bar{z}v}+{\phi}_{z\bar{v}}+{\phi}_{\bar{z}\bar{v}})e^{-a}-({\phi}_{vz}-{\phi}_{\bar{v}z}+{\phi}_{v\bar{z}}-{\phi}_{\bar{v}\bar{z}})\frac{1}{A}+\frac{(A_z+A_{\bar{z}})}{A^2}({\phi}_v-{\phi}_{\bar{v}})]\nonumber\\
& &
+e^{1456}[({\phi}_{zv}-{\phi}_{\bar{z}v}-{\phi}_{z\bar{v}}+{\phi}_{\bar{z}\bar{v}})e^{-a}-({\phi}_{vz}+{\phi}_{\bar{v}z}+{\phi}_{v\bar{z}}+{\phi}_{\bar{v}\bar{z}})\frac{1}{A}+\frac{(A_z+A_{\bar{z}})}{A^2}({\phi}_v+{\phi}_{\bar{v}})]\nonumber\\
& &
+e^{2356}[({\phi}_{zv}+{\phi}_{\bar{z}v}+{\phi}_{z\bar{v}}+{\phi}_{\bar{z}\bar{v}})e^{-a}-({\phi}_{vz}-{\phi}_{\bar{v}z}-{\phi}_{v\bar{z}}-{\phi}_{\bar{v}\bar{z}})\frac{1}{A}+\frac{(A_z-A_{\bar{z}})}{A^2}({\phi}_v-{\phi}_{\bar{v}})]\nonumber\\
& &
+ie^{2456}[({\phi}_{zv}+{\phi}_{\bar{z}v}-{\phi}_{z\bar{v}}-{\phi}_{\bar{z}\bar{v}})e^{-a}-({\phi}_{vz}+{\phi}_{\bar{v}z}-{\phi}_{v\bar{z}}-{\phi}_{\bar{v}\bar{z}})\frac{1}{A}+\frac{(A_z-A_{\bar{z}})}{A^2}({\phi}_v+{\phi}_{\bar{v}})]\nonumber\\
& &
+e^{3456}[-4{\phi}_{\bar{v}v}+\frac{2A_v{\phi}_{\bar{v}}+2A_{\bar{z}}{\phi}_v}{A^2}]\nonumber\\
& &e^{1234}[-A4{\phi}_{\bar{z}z}e^{-a}+
e^a[-4{\phi}_{\bar{v}v}+\frac{2A_v{\phi}_{\bar{v}}+2A_{\bar{z}}{\phi}_v}{A^2}]]
\end{eqnarray}
\begin{eqnarray}
& &   d{\star}_7 S =
 -\frac{1}{2} dt \wedge
[e^{1256}[2(A_{zz}+A_{\bar{z}\bar{z}})-A(\phi_{zz}+\phi_{{\bar{z}}\bar{z}})-A_z
\phi_z -A_{\bar{z}}\phi_{\bar{z}}]\nonumber\\
& &
-e^{2356}[A_{zv}+A_{\bar{z}\bar{v}}+A_{z\bar{v}}+A_{\bar{z}\bar{v}}-\frac{A}{2}(\phi_{zv}+\phi_{\bar{z}v}+\phi_{z\bar{v}}+\phi_{\bar{z}\bar{v}})\nonumber\\
&
&-\frac{(A_v+A_{\bar{v}})}{2}({\phi}_z+{\phi}_{\bar{z}})+\frac{e^a}{2}(\phi_{vz}-\phi_{\bar{v}z}-\phi_{v\bar{z}}+\phi_{\bar{v}\bar{z}})]\nonumber\\
&
&-ie^{2456}[A_{zv}+A_{\bar{z}\bar{v}}-A_{z\bar{v}}-A_{\bar{z}\bar{v}}-\frac{A}{2}(\phi_{zv}+\phi_{\bar{z}v}-\phi_{z\bar{v}}-\phi_{\bar{z}\bar{v}})\nonumber\\
&
&-\frac{(A_v-A_{\bar{v}})}{2}({\phi}_z+{\phi}_{\bar{z}})+\frac{e^a}{2}(\phi_{vz}+\phi_{\bar{v}z}-\phi_{v\bar{z}}-\phi_{\bar{v}\bar{z}})]\nonumber\\
&
&-ie^{1356}[A_{zv}-A_{\bar{z}\bar{v}}+A_{z\bar{v}}-A_{\bar{z}\bar{v}}-\frac{A}{2}(\phi_{zv}-\phi_{\bar{z}v}+\phi_{z\bar{v}}-\phi_{\bar{z}\bar{v}})\nonumber\\
&
&-\frac{(A_v+A_{\bar{v}})}{2}({\phi}_z-{\phi}_{\bar{z}})+\frac{e^a}{2}(\phi_{vz}-\phi_{\bar{v}z}+\phi_{v\bar{z}}-\phi_{\bar{v}\bar{z}})]\nonumber\\
& &
+e^{1456}[A_{zv}-A_{\bar{z}\bar{v}}-A_{z\bar{v}}+A_{\bar{z}\bar{v}}-\frac{A}{2}(\phi_{zv}-\phi_{\bar{z}v}-\phi_{z\bar{v}}+\phi_{\bar{z}\bar{v}})\nonumber\\
&
&-\frac{(A_v-A_{\bar{v}})}{2}({\phi}_z-{\phi}_{\bar{z}})+\frac{e^a}{2}(\phi_{vz}+\phi_{\bar{v}z}+\phi_{v\bar{z}}+\phi_{\bar{v}\bar{z}})]\nonumber\\
&
&+e^{1234}[2(B_{zz}+B_{\bar{z}\bar{z}}+B_{vv}+B_{\bar{v}\bar{v}})-B({\phi}_{zz}+{\phi}_{\bar{z}\bar{z}}+\phi_{vv}+\phi_{\bar{v}\bar{v}})-B_z\phi_z
-B_{\bar{z}}\phi_{\bar{z}} \nonumber\\
& &-B_v\phi_v -B_{\bar{v}}\phi_{\bar{v}}+ 4B]
-e^{3456}e^a({\phi}_{vv}+{\phi}_{\bar{v}\bar{v}})\nonumber\\
& & +
(e^{1235}-e^{1246})[B_z+B_{\bar{z}}-A_z-A_{\bar{z}}-\frac{A}{2}(\phi_z+\phi_{\bar{z}})]\nonumber\\
& & +
i(e^{2345}+e^{1346})[B_v-B_{\bar{v}}-\frac{e^a}{2}(\phi_v-\phi_{\bar{v}})]\nonumber\\
& &
-i(e^{1245}+e^{1236})[B_z-B_{\bar{z}}+A_z-A_{\bar{z}}-\frac{A}{2}(\phi_z-\phi_{\bar{z}})]\nonumber\\
& & + (-e^{1345}+e^{2346})[B_v+B_{\bar{v}}+\frac{e^a}{2}(\phi_v+\phi_{\bar{v}})]]
+\end{eqnarray}

From the above, one can calculate $*_{11}dG$ as follows:
\begin{eqnarray}
{{\star}_{11}}dG &=& e^{4\Delta}dx^0\wedge dx^1\wedge dx^2\wedge
dx^3 \wedge \nonumber\\
& &\biggl[
(\ddot{a}\frac{e^{-\phi}}{2}e^a A 
 +\frac{9}{8}e^{\phi}[Ae^{-a}(4{\phi}_z{\phi}_{\bar{z}}) +
\frac{e^a}{A}(4{\phi}_v{\phi}_{\bar{v}})]\nonumber\\
&-&\frac{3}{2}e^{\phi} [e^{-a}[-A_z{\phi}_{\bar{z}}-A_{\bar{z}}{\phi}_z+
A{\phi}_z{\phi}_{\bar{z}}] +\frac{e^a}{A}{\phi}_{v}{\phi}_{\bar{v}}] \nonumber\\
&+& \frac{3}{2}e^{\phi}[-A2{\phi}_{\bar{z}z}e^{-a}+
e^a[-2{\phi}_{\bar{v}v}+\frac{A_v{\phi}_{\bar{v}}+A_{\bar{z}}{\phi}_v}{A^2}]]\nonumber\\
&-&\frac{1}{2}[2(B_{zz}+B_{\bar{z}\bar{z}}+B_{vv}+B_{\bar{v}\bar{v}})-B({\phi}_{zz}+{\phi}_{\bar{z}\bar{z}}+\phi_{vv}+\phi_{\bar{v}\bar{v}})-B_z\phi_z
-B_{\bar{z}}\phi_{\bar{z}} \nonumber\\
& &-B_v\phi_v -B_{\bar{v}}\phi_{\bar{v}}+ 4B]-{1\over2}\dot{a}\dot{\phi}
e^{-\phi}e^aA)e^{56}\nonumber\\
&+& (\ddot{a}\frac{e^{-\phi}}{2}e^a B 
+\frac{9}{8}e^{\phi}Be^{-a}4{\phi}_z{\phi}_{\bar{z}}
-\frac{3}{2}e^{\phi}e^{-a}[-B_z{\phi}_{\bar{z}}-B_{\bar{z}}{\phi}_z+
B{\phi}_z{\phi}_{\bar{z}}] \nonumber\\
&+& 3e^{\phi}B[-{\phi}_{\bar{z}z}e^{-a}]
-\frac{1}{2}[2(A_{zz}+A_{\bar{z}\bar{z}})-A(\phi_{zz}+\phi_{{\bar{z}}\bar{z}})-A_z
\phi_z -A_{\bar{z}}\phi_{\bar{z}}]-{1\over2}\dot{a}\dot{\phi}
e^{-\phi}e^aB)e^{34} \nonumber\\
&+& (\ddot{a}\frac{e^{-\phi}}{2}AB 
 +\frac{9}{8}e^{\phi}\frac{B}{A}
4{\phi}_v{\phi}_{\bar{v}}]-\frac{3}{2}e^{\phi}\frac{1}{A}[-B_v{\phi}_{\bar{v}}-B_{\bar{v}}{\phi}_v+
B{\phi}_v{\phi}_{\bar{v}}] \nonumber\\
&+&
\frac{3}{2}e^{\phi}[-2{\phi}_{\bar{v}v}+\frac{A_v{\phi}_{\bar{v}}+A_{\bar{z}}{\phi}_v}{A^2}]
+\frac{1}{2}e^a({\phi}_{vv}+{\phi}_{\bar{v}\bar{v}})-{1\over2}\dot{a}\dot{\phi}
e^{-\phi}e^aAB)e^{12}\nonumber\\
&-& \frac{\dot{a}}{2}e^{-\phi} [(\frac{\partial}{\partial z}+
\frac{\partial}{\partial \bar{z}}) AB] e^{2t}
- \frac{\dot{a}}{2}e^{-\phi} [(\frac{\partial}{\partial z}-
\frac{\partial}{\partial \bar{z}}) AB]ie^{1t}\nonumber\\
&+&\frac{\dot{a}}{4}e^{-\phi}e^a(\phi_v+\phi_{\bar{v}})Be^{4t}
+\frac{\dot{a}}{4} e^{-\phi}e^a(\phi_v-\phi_{\bar{v}})Be^{3t}\nonumber\\
&+& (-\frac{9}{8}e^{\phi}B[-({\phi}_z-{\phi}_{\bar{z}})({\phi}_v+{\phi}_{\bar{v}})
e^{-a} + \frac{1}{A}({\phi}_z+{\phi}_{\bar{z}})({\phi}_v-{\phi}_{\bar{v}})]\nonumber\\ 
&-&\frac{3}{4}e^{\phi}[\frac{({\phi}_v
-{\phi}_{\bar{v}})}{A}[(B_z+B_{\bar{z}})-\frac{B}{2} 
({\phi}_z +{\phi}_{\bar{z}})]\nonumber\\
& &-e^{-a}({\phi}_z -{\phi}_{\bar{z}})
[(B_v+B_{\bar{v}})-\frac{B}{2} ({\phi}_v +{\phi}_{\bar{v}})]] \nonumber\\
&-& \frac{3}{4}e^{\phi}
[({\phi}_{zv}-{\phi}_{\bar{z}v}+{\phi}_{z\bar{v}}+{\phi}_{\bar{z}\bar{v}})e^{-a} -
({\phi}_{vz}-{\phi}_{\bar{v}z}+{\phi}_{v\bar{z}}-{\phi}_{\bar{v}\bar{z}})\frac{1}{A}\nonumber\\
&+&\frac{(A_z+A_{\bar{z}})}{A^2}({\phi}_v-{\phi}_{\bar{v}})]
+\frac{1}{2}[A_{zv}-A_{\bar{z}\bar{v}}+A_{z\bar{v}}-A_{\bar{z}\bar{v}}-\frac{A}{2}(\phi_{zv}-\phi_{\bar{z}v}+\phi_{z\bar{v}}-\phi_{\bar{z}\bar{v}})\nonumber\\
&-&\frac{(A_v+A_{\bar{v}})}{2}({\phi}_z-{\phi}_{\bar{z}})+\frac{e^a}{2}(\phi_{vz}-\phi_{\bar{v}z}+\phi_{v\bar{z}}-\phi_{\bar{v}\bar{z}})])ie^{24}
\nonumber\\
&+& (-\frac{9}{8}e^{\phi}B[({\phi}_z-{\phi}_{\bar{z}})
({\phi}_v-{\phi}_{\bar{v}})e^{-a} 
- \frac{1}{A}({\phi}_z+{\phi}_{\bar{z}})({\phi}_v+{\phi}_{\bar{v}})]\nonumber\\
&+&\frac{3}{4}e^{\phi}[\frac{({\phi}_v +{\phi}_{\bar{v}})}{A}[(B_z+B_{\bar{z}})
-\frac{B}{2} ({\phi}_z +{\phi}_{\bar{z}})]\nonumber\\
&-&e^{-a}({\phi}_z -{\phi}_{\bar{z}})[(B_v-B_{\bar{v}})
-\frac{B}{2} ({\phi}_v -{\phi}_{\bar{v}})]] \nonumber\\
&+& \frac{3}{4}e^{\phi}
[({\phi}_{zv}-{\phi}_{\bar{z}v}-{\phi}_{z\bar{v}}+{\phi}_{\bar{z}\bar{v}})e^{-a}-({\phi}_vz+{\phi}_{\bar{v}z}+{\phi}_{v\bar{z}}+{\phi}_{\bar{v}\bar{z}})\frac{1}{A}\nonumber\\
&+&\frac{(A_z+A_{\bar{z}})}{A^2}({\phi}_v+{\phi}_{\bar{v}})]
-\frac{1}{2}[A_{zv}-A_{\bar{z}\bar{v}}-A_{z\bar{v}}+A_{\bar{z}\bar{v}}-\frac{A}{2}(\phi_{zv}-\phi_{\bar{z}v}-\phi_{z\bar{v}}+\phi_{\bar{z}\bar{v}})
\nonumber\\
&-&\frac{(A_v-A_{\bar{v}})}{2}({\phi}_z-{\phi}_{\bar{z}})+\frac{e^a}{2}(\phi_{vz}+\phi_{\bar{v}z}+\phi_{v\bar{z}}+\phi_{\bar{v}\bar{z}})])e^{23}\nonumber\\
&+& (-\frac{9}{8}e^{\phi}B[({\phi}_z+{\phi}_{\bar{z}})
({\phi}_v+{\phi}_{\bar{v}})e^{-a} 
- \frac{1}{A}({\phi}_z-{\phi}_{\bar{z}})({\phi}_v-{\phi}_{\bar{v}})] \nonumber\\
&+&\frac{3}{4}e^{\phi}[\frac{({\phi}_v -{\phi}_{\bar{v}})}{A}
[(B_z-B_{\bar{z}})-\frac{B}{2} ({\phi}_z -{\phi}_{\bar{z}})]\nonumber\\
&-&e^{-a}({\phi}_z +{\phi}_{\bar{z}})[(B_v+B_{\bar{v}})-\frac{B}{2} ({\phi}_v
+{\phi}_{\bar{v}})]] \nonumber\\
&+&
\frac{3}{4}e^{\phi}[({\phi}_{zv}+{\phi}_{\bar{z}v}+{\phi}_{z\bar{v}}+{\phi}_{\bar{z}\bar{v}})e^{-a}-({\phi}_vz-{\phi}_{\bar{v}z}-{\phi}_{v\bar{z}}-{\phi}_{\bar{v}\bar{z}})\frac{1}{A}\nonumber\\
&+&\frac{(A_z-A_{\bar{z}})}{A^2}({\phi}_v-{\phi}_{\bar{v}})]
 +\frac{1}{2}[A_{zv}+A_{\bar{z}\bar{v}}+A_{z\bar{v}}+A_{\bar{z}\bar{v}}-\frac{A}{2}(\phi_{zv}+\phi_{\bar{z}v}+\phi_{z\bar{v}}+\phi_{\bar{z}\bar{v}})\nonumber\\
&-&\frac{(A_v+A_{\bar{v}})}{2}({\phi}_z-{\phi}_{\bar{z}})+\frac{e^a}{2}(\phi_{vz}-\phi_{\bar{v}z}-\phi_{v\bar{z}}+\phi_{\bar{v}\bar{z}})])e^{14}
\nonumber\\
&+&
(-\frac{9}{8}e^{\phi}B[({\phi}_z+{\phi}_{\bar{z}})({\phi}_v-{\phi}_{\bar{v}})e^{-a}
- 
\frac{1}{A}({\phi}_z-{\phi}_{\bar{z}})({\phi}_v+{\phi}_{\bar{v}})] \nonumber\\
&+&\frac{3}{4}e^{\phi}[\frac{({\phi}_v +{\phi}_{\bar{v}})}{A}
[(B_z-B_{\bar{z}})-\frac{B}{2} ({\phi}_z -{\phi}_{\bar{z}})]\nonumber\\
&-&e^{-a}({\phi}_z +{\phi}_{\bar{z}})[(B_v-B_{\bar{v}})-\frac{B}{2} 
({\phi}_v -{\phi}_{\bar{v}})]]\nonumber\\
&+&
\frac{3}{4}e^{\phi}[({\phi}_{zv}+{\phi}_{\bar{z}v}-{\phi}_{z\bar{v}}-{\phi}_{\bar{z}\bar{v}})e^{-a}-({\phi}_vz+{\phi}_{\bar{v}z}-{\phi}_{v\bar{z}}-{\phi}_{\bar{v}\bar{z}})\frac{1}{A}\nonumber\\
&+&\frac{(A_z-A_{\bar{z}})}{A^2}({\phi}_v+{\phi}_{\bar{v}})]
+\frac{1}{2}[A_{zv}+A_{\bar{z}\bar{v}}-A_{z\bar{v}}-A_{\bar{z}\bar{v}}-\frac{A}{2}(\phi_{zv}+\phi_{\bar{z}v}-\phi_{z\bar{v}}-\phi_{\bar{z}\bar{v}})\nonumber\\
&-&\frac{(A_v-A_{\bar{v}})}{2}({\phi}_z+{\phi}_{\bar{z}})+\frac{e^a}{2}(\phi_{vz}+\phi_{\bar{v}z}-\phi_{v\bar{z}}-\phi_{\bar{v}\bar{z}})])ie^{13}\nonumber\\
&+& (\frac{3}{4}e^{\phi}e^{-a}B({\phi}_z -{\phi}_{\bar{z}}) 
+\frac{1}{2}[B_z-B_{\bar{z}}+A_z-A_{\bar{z}}-\frac{A}{2}(\phi_z-\phi_{\bar{z}})])i(e^{45}+e^{36})\nonumber\\
&+&(\frac{3}{4}e^{\phi}e^{-a}B({\phi}_z +{\phi}_{\bar{z}}) 
+\frac{1}{2}[B_z+B_{\bar{z}}-A_z-A_{\bar{z}}+\frac{A}{2}(\phi_z+\phi_{\bar{z}})])(e^{35}-e^{46})\nonumber\\
&+&(-\frac{3}{4}e^{\phi}\frac{({\phi}_v -{\phi}_{\bar{v}})}{A} B
-\frac{1}{2}[B_v-B_{\bar{v}}-\frac{e^a}{2}(\phi_v-\phi_{\bar{v}})])i(e^{25}+e^{16})\nonumber\\
&+&(-\frac{3}{4}e^{\phi}\frac{({\phi}_v +{\phi}_{\bar{v}})}{A}B
 -\frac{1}{2}[B_v+B_{\bar{v}}+\frac{e^a}{2}(\phi_v+\phi_{\bar{v}})])(e^{15}-e^{26})
\biggr] \nonumber\\
\end{eqnarray}
This is relevant to the interpretation of the uplift as $M5$-branes
wrapped around supersymmetric two-cycles calibrated by a generalized
calibration $J$, but generically, away from the standard complex structure
point in the moduli space of almost complex structures. In the absence of 
$M5$-branes, $*_{11}dG=0$ would lead to a system of coupled non-linear
partial differential equations which will, if the solutions exists,  be
very difficult to solve.

\section{The conformal immersion  in ${\bf C}^3$}
\setcounter{equation}{0}
\seceqbb

In this appendix, we give an example of a Riemann surface obtained as a
Weierstra$\ss$
representation of conformal immersion of a surface in ${\bf C}^3$,
using $\phi=x\equiv$ Re(z) (the ``linear dilaton" ansatz) and
$A,B\in{\bf R}$. As the solutions we get will not involve periodic
$A,B,\phi$, they are relevant only locally to the (non-compact) seven-folds of
$SU(3)$-structure. 

One needs to solve the following differential equation
(use ${\partial A,B\over\partial y}=0$)
\begin{equation}
({1\over 4}{d^2\over dx^2} - {1\over 2x}{d\over dx} + x^2)B=0.
\end{equation}
The two solutions, $B_{1,2}$ are given by:
\begin{equation}
\label{eq:B1B2}
B_1(x)=x^{{3\over 2}}J_{{3\over 4}}(x^2),\ B_2(x)=x^{{3\over2}}J_{-{3\over 4}}
(x^2).
\end{equation}
Hence, the corresponding two values for $A$ would be:
\begin{eqnarray}
\label{eq:A1A2}
& & A_1(x)={1\over2\sqrt{x}}(-2x^2J_{-{1\over4}}(x^2) -3J_{{3\over4}}(x^2) + 
2x^2J_{{7\over4}}(x^2)),\nonumber\\
& & A_2(x)={1\over2\sqrt{x}}(-2x^2J_{-{7\over4}}(x^2) -3J_{-{3\over4}}(x^2) + 
2x^2J_{{1\over4}}(x^2)).
\end{eqnarray}
Then the Weierstra$\ss$ representation for conformal immersion of a surface
($\equiv$ Riemann surface) in ${\bf C}^3$ (See \cite{confimrs} and references
therein) is given by defining the following coordinates $X^{i=1,...,6}(z,
{\bar z})$, of the Rieman surface in ${\bf C}^3$:
\begin{eqnarray}
\label{eq:immers}
& & X^1 + i X^2 = i\int_\Gamma(A_1^2 dz^\prime - B_1^2 d{\bar z}^\prime),
\nonumber\\
& & X^1 - i X^2 = i\int_\Gamma(B_1^2dz^\prime - A_1^2d{\bar z}^\prime),
\nonumber\\
& & X^3 =  -\int_\Gamma(A_1B_1dz^\prime + A_1B_1d{\bar z}^\prime),
\nonumber\\
& & X^4 + i X^5 = i\int_\Gamma(A_2^2 dz^\prime - B_2^2 d{\bar z}^\prime),
\nonumber\\
& & X^4 - i X^5 = i\int_\Gamma(B_2^2dz^\prime - A_2^2d{\bar z}^\prime),
\nonumber\\
& & X^6 =  -\int_\Gamma(A_2B_2dz^\prime + A_2B_2d{\bar z}^\prime),
\end{eqnarray} 
where $\Gamma$ is a contour in ${\bf C}(z,{\bar z})$. The ${\bf C}^3$ 
coordinates for values of $B_{1,2}$ and $A_{1,2}$ as given in
(\ref{eq:B1B2}) and (\ref{eq:A1A2}), for the path
$\Gamma=\{(x,y)|x=y\}$, is given in appendix B. The induced metric
on the Riemann surface is then given by:
\begin{equation}
ds^2=(A_1^2 + A_2^2 + B_1^2 + B_2^2)^2|dz|^2.
\end{equation}
The condition for a conformal immersion 
(implying $g_{zz}=g_{{\bar z}{\bar z}}=0$):
\begin{equation}
({\partial X^1\over\partial z})^2 + 
({\partial X^2\over\partial z})^2 + 
({\partial X^3\over\partial z})^2 +
({\partial X^4\over\partial z})^2 +
({\partial X^5\over\partial z})^2 +
({\partial X^6\over\partial z})^2 = 0,
\end{equation}
is a quadric in ${\bf CP}^5$:
\begin{equation}
w_1^2+w_2^2+w_3^2+w_4^2+w_5^2+w_6^2=0,
\end{equation}
$w_i$ being homogeneous coordinates on ${\bf CP}^5$.

Along the path $\Gamma:\{(x^\prime,y^\prime)|y^\prime=x^\prime, 0\leq
x^\prime\leq x\neq0,x<1\}$, one sees that:
\begin{equation}
X^1 + i X^2 = iI_2 - I_1,\ X^3 = -I_3;\ X^4 + i X^5 = iI_5 - I_4,\ X^6 = -I_6,
\end{equation}
where 
\begin{eqnarray}
& & I_1\equiv\int_\Gamma(A_1^2 + B_1^2),\ I_2\equiv\int_\Gamma(A_1^2 - B_1^2),\
I_3\equiv\int_\Gamma A_1B_1
;\nonumber\\
& & I4_\equiv\int_\Gamma(A_1^2 + B_1^2),\ I_5\equiv\int_\Gamma(A_1^2 - B_1^2),\
I_6\equiv\int_\Gamma A_1B_1.
\end{eqnarray}
The integrals $I_i, i=1,...,6$ are given as:
\begin{eqnarray}
& & 
I_1=\frac{x^4\,\left((J_{-\frac{1}{4}}(x^2))^2 -J_{-\frac{5}{4}}(x^2)\,
J_{\frac{3}{4}}(x^2)\right)}{4} + 
  \frac{x^4\, \left((J_{\frac{3}{4}}(x^2))^2 -J_{-\frac{1}{4}}(x^2)\,
J_{\frac{7}{4}}(x^2)\right)}{4} \nonumber\\
& & + \frac{x^4\, \left((J_{\frac{7}{4}}(x^2))^2 - 
J_{\frac{3}{4}}(x^2)\, J_{\frac{11}{4}}(x^2)\right)}{4} +
\frac{{\left( x^2 \right) }^{\frac{3}{2}}\,\ _2F_3(\{ \frac{3}{4},
       \frac{5}{4}\} , \{ \frac{3}{2}, \frac{7}{4}, \frac{7}{4}\} ,-x^4)}
     {{\sqrt{2}}\,\Gamma(\frac{3}{4})\, \Gamma(\frac{7}{4})} + \nonumber\\
& &   \frac{3\,{\left( x^2 \right)}^{\frac{3}{2}}\, 
\ _2F_3(\{ \frac{3}{4}, \frac{5}{4}\} , \{ \frac{7}{4}, 
\frac{7}{4}, \frac{5}{2}\} ,-x^4)} 
{8\,{\sqrt{2}}\, \Gamma(\frac{7}{4})}^2 
- \frac{3\,{\left( x^2 \right)}^{\frac{7}{2}}
\, \ _2F_3(\{ \frac{7}{4}, \frac{9}{4}\}, 
\{ \frac{11}{4}, \frac{11}{4}, \frac{7}{2}\} ,-x^4)} {28\,{\sqrt{2}}\,
\Gamma(\frac{7}{4})\, \Gamma(\frac{11}{4})}\nonumber\\
& & -\frac{{\left(x^2\right)}^{\frac{7}{2}}\,\left( 363\,
\ _2F_3(\{ \frac{5} {4},\frac{7}{4}\} , \{ \frac{5}{2}, \frac{11}{4},
 \frac{11}{4}\} , -x^4) - 56\,x^4\,
\ _2F_3(\{ \frac{9}{4},\frac{11}{4}\},\{ \frac{7}{2}, 
\frac{15}{4},\frac{15}{4}\} , -x^4) \right) }{2541\,{\sqrt{2}}\,
\Gamma(\frac{3}{4})\,\Gamma(\frac{11}{4})}
\end{eqnarray}

\begin{eqnarray}
& & I_2=\frac{x^4\,\left((J_{-\frac{1}{4}}(x^2))^2 - J_{-\frac{5}{4}}(x^2)\,
J_{\frac{3}{4}}(x^2)\right) }{4} - \frac{x^4\,
     \left((J_{\frac{3}{4}}(x^2))^2 - J_{-\frac{1}{4}}(x^2)\, 
J_{\frac{7}{4}}(x^2)\right) }{4} \nonumber\\
& & + \frac{x^4\,
\left((J_{\frac{7}{4}}(x^2))^2 - J_{\frac{3}{4}}(x^2)\, J_{\frac{11}{4}}(x^2)
\right) }{4} + \frac{{\left( x^2 \right) }^{\frac{3}{2}}\,
\ _2F_3(\{ \frac{3}{4}, \frac{5}{4}\} ,\{ \frac{3}{2}, \frac{7}{4},
\frac{7}{4}\} ,-x^4)}{{\sqrt{2}}\, \Gamma(\frac{3}{4})\,
\Gamma(\frac{7}{4})} + \nonumber\\
& & \frac{3\,{\left( x^2 \right)}^{\frac{3}{2}}\,
\ _2F_3(\{ \frac{3}{4},\frac{5}{4}\} , \{ \frac{7}{4},\frac{7}{4},
\frac{5}{2}\} ,-x^4)}{8\,{\sqrt{2}}\,(\Gamma(\frac{7}{4}))^2} - 
  \frac{3\,{\left( x^2 \right)}^{\frac{7}{2}}\,
\ _2F_3(\{ \frac{7}{4}, \frac{9}{4}\} , \{ \frac{11}{4}, \frac{11}{4},
\frac{7}{2}\} ,-x^4)}{28\,{\sqrt{2}}\, \Gamma(\frac{7}{4})\,\Gamma(
\frac{11}{4})}\nonumber\\
& &  - \frac{{\left( x^2 \right) }^{\frac{7}{2}}\,\left( 363\,
\ _2F_3(\{ \frac{5}{4},\frac{7}{4}\} ,\{ \frac{5}{2},\frac{11}{4},
\frac{11}{4}\} ,-x^4) - 56\,x^4\, \ _2F_3(\{
\frac{9}{4},\frac{11}{4}\},
\{ \frac{7}{2},\frac{15}{4},\frac{15}{4}\} ,-x^4) \right) }{2541\,{\sqrt{2}}\,
\Gamma(\frac{3}{4})\, \Gamma(\frac{11}{4})}
\end{eqnarray}

\begin{eqnarray}
& & I_3=\frac{x^4}
{180\, {\sqrt{2}}\,\Gamma(\frac{3}{4})\, {\Gamma(\frac{7}{4})}^2\,\Gamma
(\frac{11}{4})}
\biggl[ -36\,\sqrt{x^2}\Gamma(\frac{7}{4})\,
\Gamma(\frac{11}{4})\, 
\ _2F_3\biggl(\{ \frac{5}{4},
\frac{5}{4}\}, \{\frac{3}{2},\frac{7}{4},\frac{9}{4}\},
-x^4\biggr)\nonumber\\
& &  + \Gamma(\frac{3}{4})\, \biggl( -27\,\Gamma(\frac{11}{4})\,
\ _2F_3\biggl(\{ \frac{5}{4},\frac{5}{4}\},
\{\frac{7}{4},\frac{9}{4},\frac{5}{2}\},-x^4\biggr)+5\,x^4\,
\Gamma(\frac{7}{4})\, \ _2F_3\biggl(\{\frac{9}{4},\frac{9}{4}\} ,
\{ \frac{11}{4},\frac{13}{4},\frac{7}{2}\} ,-x^4\biggr)\biggr))\biggr]
\nonumber\\
& & 
\end{eqnarray}

\begin{eqnarray}
& & I_4=\frac{x^4\,\left( (J_{-\frac{7}{4}}(x^2))^2 
- J_{-\frac{11}{4}}(x^2)\, J_{-\frac{3} {4}}(x^2)\right) }{4} 
+ \frac{x^4\, (\left(J_{-\frac{3}{4}}(x^2))^2 - J_{-\frac{7}{4}}(x^2)\,
J_{\frac{1}{4}}(x^2) \right) }{4} + \nonumber\\
& & \frac{x^4\,\left((J_{\frac{1}{4}}(x^2))^2-
J_{-\frac{3}{4}}(x^2)\, J_{\frac{5}{4}}(x^2)\right)}{4} -\
\frac{4\,{\sqrt{2}}\, \ _2F_3(\{ -\frac{3}{4} ,
-\left( \frac{1} {4} \right) \} , \{ - \frac{3}{2},\frac{1}{4},
\frac{1}{4}\} ,-x^4)} {{\left( x^2 \right) }^{\frac{3}{2}}\, 
\Gamma(-\frac{3}{4}),\Gamma(\frac{1}{4})} \nonumber\\
& & - \frac{3\, \ _2F_3
(\{ -\left( \frac{3}{4} \right) ,-\left(\frac{1}{4}\right)\},
\{-\left(\frac{1}{2}\right),\frac{1}{4},\frac{1}{4}\},-x^4)}{{\sqrt{2}}\,
{\left( x^2 \right) }^{\frac{3}{2}}\,(\Gamma(\frac{1}{4})^2} - 
 \frac{3\,{\sqrt{2}}\, {\sqrt{x^2}}\,
\ _2F_3(\{\frac{1}{4},\frac{3}{4}\},\{ \frac{1}{2},\frac{5}{4},\frac{5}{4}\},
-x^4)} {\Gamma( \frac{1}{4})\, \Gamma( \frac{5}{4})} \nonumber\\
& & - \frac{4\,{\sqrt{2}}\, {\sqrt{x^2}}\, \left( 75\, 
\ _2F_3(\{ -\frac{1}{4},\frac{1}{4}\},
\{ -\left(\frac{1}{2}\right),\frac{5}{4}, \frac{5}{4}\} ,-x^4) + 8\,
x^4\,\ _2F_3(\{ \frac{3} {4},\frac{5}{4}\} , \{ \frac{1}{2}, \frac{9}{4},
\frac{9}{4}\} ,-x^4)\right) }{75\,\Gamma(-\frac{3}{4})
\,\Gamma(\frac{5}{4})}\nonumber\\
& & 
\end{eqnarray}

\begin{eqnarray}
& & I_5=\frac{x^4\,\left((J_{-\frac{7}{4}}(x^2))^2 - 
J_{-\frac{11}{4}}(x^2)\, J_{-\frac{3}{4}}(x^2)\right) }{4} - 
 \frac{x^4\, \left((J_{-\frac{3}{4}}(x^2))^2 
- J_{-\frac{7}{4}}(x^2)\, J_{\frac{1}{4}}(x^2)\right) }{4} + \nonumber\\
& & \frac{x^4\,\left((J_{\frac{1}{4}}(x^2))^2 - J_{-\frac{3}{4}}(x^2)\,
J_{\frac{5}{4}}(x^2)\right) }{4} - \frac{4\,{\sqrt{2}}\,
\ _2F_3(\{ -\left( \frac{3}{4} \right) ,-\left( \frac{1} {4} \right) \} ,
\{ -\left( \frac{3} {2} \right) , \frac{1}{4}, \frac{1}{4}\} ,-x^4)}
{{\left( x^2 \right) }^{\frac{3}{2}}\, \Gamma(-\left(\frac{3}{4}\right))
\,\Gamma(\frac{1}{4})}\nonumber\\
& &  - \frac{3\,\ _2F_3(\{ -\frac{3}{4} 
,-\frac{1}{4}\},\{ -\frac{1}{2},\frac{1}{4},
\frac{1}{4}\} ,-x^4)} {{\sqrt{2}}\, {\left( x^2 \right) }^{\frac{3}{2}}\,
{\Gamma(\frac{1}{4})}^2} - \frac{3\,{\sqrt{2}}\, {\sqrt{x^2}}\,
\ _2F_3(\{ \frac{1}{4}, \frac{3}{4}\},\{ \frac{1}{2},\frac{5}{4},\frac{5}{4}\},
-x^4)} {\Gamma(\frac{1}{4})\, \Gamma(\frac{5}{4})} \nonumber\\
& & - \frac{4\,{\sqrt{2}}\,
 {\sqrt{x^2}}\, \left( 75\, \ _2F_3(\{ -\frac{1}{4},
\frac{1}{4}\},\{ -\left( \frac{1}{2}\right),\frac{5}{4},\frac{5}{4}\} ,-x^4)
 + 8\,x^4\, \ _2F_3(\{\frac{3} {4},\frac{5}{4}\},\{ \frac{1}{2},\frac{9}{4},
\frac{9}{4}\} ,-x^4)\right) }{75\,\Gamma(-\frac{3}{4})\,
\Gamma(\frac{5}{4})}\nonumber\\
& & 
\end{eqnarray}

\begin{eqnarray}
& & I_6=\frac{{\sqrt{2}}}
{3\, {\sqrt{x^2}}\, \Gamma(- \frac{3}{4})\ {\Gamma(\frac{1}{4})}^2\,\Gamma(
\frac{5}{4})}\biggl[ 12\, \Gamma(\frac{1}{4})\, 
\Gamma(\frac{5}{4})\, \ _2F_3\biggl(
\{-\frac{1}{4},-\frac{1}{4}\},
\{-\frac{3}{2},\frac{1}{4},\frac{3}{4}\},-x^4\biggr)
\nonumber\\
& & \hskip -.8cm + \Gamma(-\frac{3}{4})\,\biggl( 9\,
\Gamma(\frac{5}{4})\, \ _2F_3\biggl(\{ -\frac{1}{4}),
-\frac{1}{4}\},\{-\frac{1}{2},\frac{1}{4},
\frac{3}{4}\},-x^4\biggr)+x^4\, \Gamma( \frac{1}{4})\, \ _2F_3\biggl(
\{\frac{3}{4},
\frac{3}{4}\} , \{ \frac{1}{2}, \frac{5}{4}, \frac{7}{4}\},-x^4\biggr)
\biggr)\biggr].\nonumber\\
& & 
\end{eqnarray}
We thus see that the immersion is severely non-linear in nature for the
``linear dilaton background".

\end{document}